\documentclass[aps, prl, reprint, superscriptaddress, twocolumn, floatfix, showkeys, amsmath, amssymb, longbibliography]{revtex4-2}
\usepackage{amsmath,amssymb,bbm,mathrsfs,bm,braket,color,graphicx,comment,amsfonts,dsfont}
\usepackage[colorlinks,citecolor=blue,urlcolor=blue]{hyperref}
\usepackage{graphicx}
\usepackage{float}
\usepackage[normalem]{ulem}
\usepackage{comment}
\usepackage{tabularx}

\begin{document}
\title{Antiferro-Chiral Phonons as Conjugate Fields to N\'eel Order in $\mathcal{P}\mathcal{T}$-Symmetric Antiferromagnets}
\author{Sanjib Kumar Das}
\affiliation{Department of Physics and Astronomy, University of Delaware, Newark, Delaware 19716, USA}
\author{Randy Yeh}
\affiliation{Department of Physics and Astronomy, University of Delaware, Newark, Delaware 19716, USA}
\author{Eider A. Vega Ramirez}
\affiliation{Department of Physics and Astronomy, University of Delaware, Newark, Delaware 19716, USA}
\author{Yafei Ren}
\email{yfren@udel.edu}
\affiliation{Department of Physics and Astronomy, University of Delaware, Newark, Delaware 19716, USA}

\begin{abstract}
Chiral phonons act as a phononic conjugate field to net magnetization, enabling phononic detection and control of ferromagnetic order. Whether an analogous phononic coupling exists for the compensated N\'eel order of antiferromagnets has remained open.
Here we show that, in $\mathcal{P}\mathcal{T}$-symmetric antiferromagnets, phonons with sublattice-staggered angular momentum can couple to the N\'eel vector like a conjugate field. Such coupling can be induced by the molecular Berry curvature-induced gyroscopic coupling between Raman and infrared-active phonons. 
The gyroscopic coupling is odd under both $\mathcal{P}$ and $\mathcal{T}$ while preserving their product. 
$\mathcal{P}\mathcal{T}$ symmetry forbids the resulting normal modes to have a net phonon angular momentum (PAM) while allows counter-rotating local motion on inversion-related sublattices, which are dubbed the antiferro-chiral phonons (AFCPs). Reversing the N\'eel vector reverses the staggered phonon chirality and the sign of the coupling.
Our results indicate AFCPs can serve as probes and control knobs of N\'eel order.
\end{abstract}

\maketitle

Antiferromagnetic (AFM) materials offer a promising platform for spintronics because their compensated magnetic moments suppress stray fields and their order parameter can evolve on ultrafast time scales \cite{Gomonay2014,Baltz2018,elezn2018,Manchon19,Han2023,Rene2025}. The same compensation, however, makes their fundamental order parameter, the N\'eel vector, difficult to detect and control with probes that couple primarily to a net magnetization \cite{Wadley2016,Moriyama2018,Chiang2019}. 
Lattice dynamics has emerged as a powerful route to magnetic order through phonons carrying nonzero angular momentum, which we broadly refer to as chiral phonons \cite{Strohm2005,Zhang2010,Qin2012,Zhang2014,Chen2015,Zhang2015,Zhu2018,Chen2018,Chen2019,Ptok2021, Saparov2022, Chen2024, Ren2024PRX,Ishito2022,Ueda2023,Ueda2025,Yang2026,Juraschek2025,Wang2025alteraxial,Zhang2023Weylphonon,Ma2024,Weissenhofer2025,Che2025,Streib2021,Zhang2022chiralpam,Wang2024,Chen2025,Liu2017,Li2020,Park2020,Sun2021,Oh2025,Kim2023,Yang2024,Whofer2025, yaniv2025multicolor}. The angular momentum carried by phonons naturally enables their coupling to ferromagnetic order, allowing them to act as an effective conjugate field \cite{Juraschek2019, Luo2023}. 
In particular, reversing the magnetization flips the phonon angular momentum (PAM) \cite{Wu2025,Che2025,Yang2025} whereas optically excited coherent chiral phonons can drive spin dynamics and even induce magnetization switching \cite{Nova2016,Juraschek2019,Ren2021,Juraschek2022,Luo2023,Geilhufe2023,Zhang2023,Chaudhury2024,Wu2025,Yao2025}. 
An analogous coupling in which a phononic degree of freedom acts as a conjugate field to the N\'eel vector has yet to be established, limiting phonon-based detection and control of AFM order.

In this Letter, we demonstrate such a coupling between the N\'eel vector and a phononic degree of freedom, which we refer to as an \emph{antiferro-chiral phonon} (AFCP).
An AFCP is a phonon mode whose local angular momenta are opposite on inversion-related sublattices, so that the total PAM vanishes while the sublattice-staggered PAM ($\bm{L}_{\rm N}$) remains finite. Because $\bm{L}_{\rm N}$ is odd under both $\mathcal{P}$ and $\mathcal{T}$ while even under $\mathcal{P}\mathcal{T}$, it shares the same symmetry as the N\'eel vector. 
This symmetry equivalence dictates a leading-order phonon-magnetism coupling of the Landau type $H_{\rm eff}=-g N\hat{z}\cdot\bm{L}_{\rm N}$, establishing $\bm{L}_{\rm N}$ as the phononic conjugate field to the N\'eel vector $N\hat{z}$.
We show that Brillouin-zone-center AFCPs emerge from a gyroscopic hybridization between Raman-active and infrared-active (IR) optical phonons. This hybridization converts linearly polarized parity eigenmodes into antiferro-chiral modes that shows both Raman and IR characters. Coherently excited IR phonons can thereby excite modes and coherent spin precession. Microscopically, the coupling can be generated by the molecular Berry curvature that couples the Raman-IR subspace, with its sign locked to the N\'eel vector. 
We demonstrate this mechanism in a prototype lattice model showing both degenerate and nondegenerate AFCPs. Reversing the N\'eel vector reverses the staggered phonon chirality. 

\emph{Symmetry analysis.---}\label{symmetry}
$\mathcal{P}\mathcal{T}$ symmetric phonons can be characterized by the sublattice-staggered PAM $\bm{L}_{\rm N}$. To define $\bm{L}_{\rm N}$, we partition the atoms into two groups, $A$ and $B$, exchanged by the inversion operation $\mathcal{P}$, and define the angular momentum on each group as $\bm{L}_A=\sum_{s\in A}\bm{l}_s$ and $\bm{L}_B=\sum_{s\in B}\bm{l}_s$ with $\bm l_s$ being the angular momentum of atom $s$ \cite{Zhang2014, Coh2023,Wang2025alteraxial, Ren2024PRX}. $\mathcal{P}\mathcal{T}$ invariance makes $\bm{L}_A=-\bm{L}_B$ because $\mathcal{P}$ exchanges $A$ and $B$ while $\mathcal{T}$ reverses angular momentum. Therefore, the total PAM $\bm{L}=\bm{L}_A+\bm{L}_B$ vanishes while the sublattice-staggered PAM $\bm{L}_{\rm N}=\bm{L}_A-\bm{L}_B$ can remain finite. This qualitative conclusion is independent of the specific partition, as long as the two groups are exchanged by inversion.

A nonzero local PAM requires a phase difference between displacement components. The angular momentum of site $s$ is $\bm{l}_s = M_s\,\mathrm{Im}(\bm{e}_{\nu s}^* \times \bm{e}_{\nu s})$, where $\bm{e}_{\nu s}$ is the polarization vector of phonon mode $\nu$ on that site and $M_s$ is the ionic mass \cite{Zhang2014}. To have a finite $\bm{l}_s$, the polarization vector must not be choosable as purely real.
At a generic nonzero wave vector, the dynamical matrix and phonon polarization vectors are generally complex, leading to phonon eigenmodes with atom-resolved angular momentum. If the lattice potential energy is inversion symmetric, the nondegenerate phonon modes must have zero net angular momentum, but can have nonzero sublattice-staggered angular momentum, such as the phonons at the Brillouin-zone corner $K$ in an inversion-symmetric honeycomb lattice~\cite{Zhang2014, chaudhary2026anomalous}. Breaking the inversion symmetry allows the phonons to carry nonzero net angular momentum and nondegenerate chiral phonons~\cite{Zhang2014}. Based on the symmetry constrained momentum-space texture of the net PAM and local PAM pattern, a recent classification distinguishes ferroaxial, antiferro-nonaxial, and alteraxial phonons~\cite{Wang2025alteraxial}. The AFCPs considered here is the same as the antiferro-nonaxial phonon sector: their opposite local rotations produce a finite sublattice-staggered PAM $\bm L_{\rm N}$ and zero net PAM.

In contrast to the symmetry classification, here we focus on the interplay between the AFCP at the Brillouin-zone center $\Gamma$ and the N\'eel vector. We use the term AFCP to emphasize that the sublattice-staggered PAM couples directly to the N\'eel vector as shown below, in analogy with a conventional chiral phonon whose net PAM couples to ferromagnetic order. At $\Gamma$, 
the dynamical matrix in a mass-spring model is real symmetric. Every nondegenerate eigenmode, as well as basis within each degenerate subspace, can be chosen real. Intrinsically complex zone-center phonons with atom-resolved PAM require a time-reversal-odd dynamical correction~\cite{Saparov2022,Bonini2023,Ren2024PRX,Wang2025ab,Wang2025alteraxial} and cannot be induced by a static spin-dependent scalar potentials originated from, e.g., the magnetoelastic coupling or phonon-dependent exchange or anisotropy energy~\cite{fennie2006magnetically,kant2012universal, sohn2017strong, son2019unconventional}.

\emph{Minimal effective model.---}\label{effmodel}
Beyond the scalar-potential description, dynamical corrections to the phonons induced by N\'eel order must transform under $\mathcal P$ and $\mathcal T$ in the same way as the underlying N\'eel vector. The $\mathcal{P}$ oddness forbids couplings within the Raman (or IR) sector and thus requires inter-parity hybridization between them. The $\mathcal{T}$ oddness requires this Raman-IR coupling to be dynamical and velocity dependent. Here, we illustrate the symmetry-constrained coupling using a two-mode minimal model.

We focus on a pair of optical phonons, $\bm{Q}=(Q_{\rm R},Q_{\rm IR})^{\rm T}$, where $Q_{\rm R}$ and $Q_{\rm IR}$ denote the coordinates of Raman- and IR-active normal modes defined by the bare scalar potential. Both modes are real and linearly polarized, carrying zero PAM. Their bare squared frequencies are $\omega_0^2\pm\Delta$, corresponding to a scalar potential described by the unperturbed dynamical matrix $D_0=\omega_0^2\sigma_0+\Delta\sigma_z$.
The leading-order correction induced by the AFM order and allowed by symmetry is a gyroscopic coupling in the Raman-IR mode space as shown in the last term in the Lagrangian
\begin{equation}
    L_{\rm eff}
    = \frac{1}{2}\dot{\bm Q}^{\,2}
    - \frac{1}{2}\bm Q^{\rm T} D_0 \bm Q
    + \frac{\Omega}{2}\,\hat z\cdot(\bm Q\times\dot{\bm Q}),
\end{equation}
where $\dot{\bm{Q}}$ is the corresponding velocity and $\hat z$ is the unit vector perpendicular to the internal $(Q_{\rm R},Q_{\rm IR})$ mode space. The last term has the form of a Lorentz coupling to an effective magnetic field $\Omega\hat z$ that corresponds to $-gN$ in the previously mentioned $H_{\rm eff}$. This term is $\mathcal{P}$ odd as the cross product mixes the opposite-parity Raman and IR coordinates, while the velocity $\dot{\bm{Q}}$ makes the coupling $\mathcal{T}$ odd. It is even under $\mathcal{P}\mathcal{T}$. 

\begin{figure}[t!]
    \centering
    \includegraphics[width=\columnwidth]{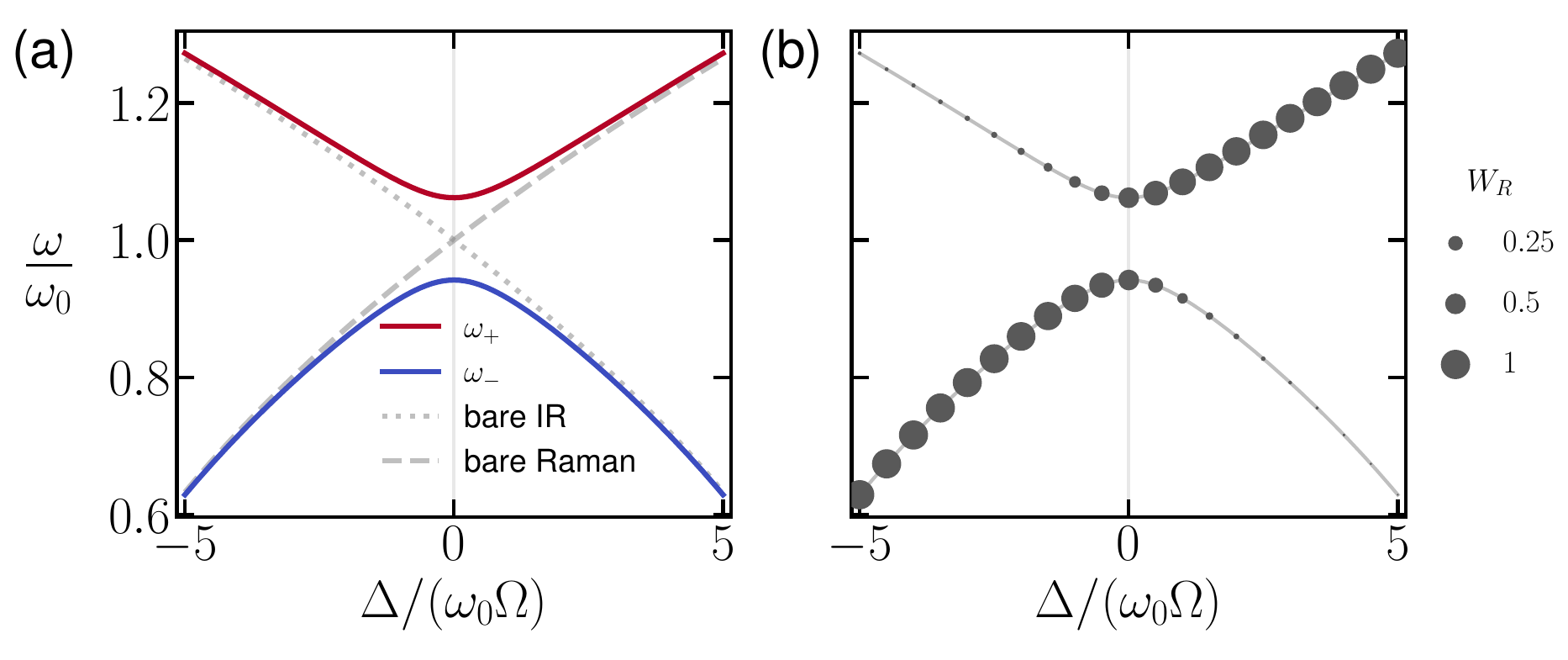}
    \caption{Two-mode gyroscopic hybridization vs detuning $\Delta/(\omega_0\Omega)$. (a) Hybrid frequencies $\omega_\pm$ (solid lines), the bare IR frequency (gray dotted line), and the bare Raman frequency (gray dashed line). (b) The same hybrid phonon branches, with the circle area representing the normalized Raman-character weight $W_{\rm R}=|Q_{\rm R}|^2/(|Q_{\rm R}|^2+|Q_{\rm IR}|^2)$.}
    \label{fig:minimal_spectrum_weight}
\end{figure}

The Lagrangian leads to an Euler-Lagrange equation for the phonon modes,
\begin{equation}
    \ddot{\bm Q}+D_0\bm Q+\Omega\,\hat z\times\dot{\bm Q}=0.
\end{equation}
Using the harmonic ansatz $\bm Q(t)=\Psi e^{-i\omega t}$, one obtains the frequency-domain mode equation
\begin{equation}\label{eq:maineffeqn}
    \left[D_0-\omega^2\sigma_0+i\omega\Omega\,\varepsilon\right]\Psi=0,
\end{equation}
where the gyroscopic coupling enters as a frequency-dependent term proportional to $i\omega\Omega$ with $\varepsilon$ the antisymmetric tensor,
\begin{equation}
    \varepsilon=
    \begin{pmatrix}
        0 & 1\\
        -1 & 0
    \end{pmatrix}
    = i\sigma_y.
\end{equation}
Solving for $\omega^2$ gives the two hybridized branches as shown in Fig.~\ref{fig:minimal_spectrum_weight}(a),
\begin{equation}
    \omega_\pm^2
    = \omega_0^2+\frac{\Omega^2}{2}
    \pm \sqrt{\Delta^2+\omega_0^2\Omega^2+\frac{\Omega^4}{4}}.
\end{equation}
The corresponding eigenvector for each branch may be chosen as
\begin{equation}
    \Psi_\pm
    = \frac{1}{\mathcal N_\pm}
    \begin{pmatrix}
        \omega_\pm\Omega\\
        i(\omega_0^2+\Delta-\omega_\pm^2)
    \end{pmatrix},
\end{equation}
with $\mathcal N_\pm$ the normalization factor. The Raman and IR components therefore differ by a relative phase of $\pi/2$, showing that the hybridized modes are no longer linearly polarized. This phase locking is the direct signature of the effective magnetic field in the internal Raman-IR mode space.

Figure~\ref{fig:minimal_spectrum_weight}(b) illustrates the redistribution of Raman character induced by the gyroscopic hybridization. Here, the area of the circle represents the normalized Raman-component weight $W_{\rm R}$ of the phonon eigenvector. At resonance, $\Delta=0$, both hybrid modes have equal Raman and IR component weights. 
As $|\Delta|$ increases, one branch becomes predominantly Raman-like while the other becomes predominantly IR-like. 
The measured Raman and IR intensities further depend on the bare-mode Raman tensors and Born effective charges.
Magnons with magnetic dipole can have IR activity. Their coupling with Raman-active phonon modes can also induce spectral weight transfer~\cite{Cui2023}, which can be distinguished from the IR-active phonons by the different magnetic-field dependences between magnons and phonons.

The gyroscopic term is directly tied to the staggered PAM. Let $\bm{e}_{{\rm R},s}$ and $\bm{e}_{{\rm IR},s}$ denote the real displacement patterns of the bare Raman and IR modes on site $s$. The instantaneous lattice displacement is then
\begin{equation}
    \bm{u}_s(t)=Q_{\rm R}(t)\bm{e}_{{\rm R},s}+Q_{\rm IR}(t)\bm{e}_{{\rm IR},s}.
\end{equation}
The local angular momentum on site $s$ is
\begin{equation}
    \bm{l}_s=M_s\,\bm{u}_s\times\dot{\bm{u}}_s
    =(\hat{z}\cdot\bm{Q}\times\dot{\bm{Q}}) M_s 
    \bigl(\bm{e}_{{\rm R},s}\times\bm{e}_{{\rm IR},s}\bigr).
\end{equation}
Summing over sublattices with the staggering factor $\eta_s=\pm1$ for $s\in A$ or $s\in B$ gives
\begin{equation}
    \bm{L}_{\rm N}
    = (\hat{z}\cdot \bm Q\times\dot{\bm Q}) \, \bm \chi_{\rm N}
\end{equation}
where $\bm \chi_{\rm N} = \sum_s \eta_s M_s\, \bigl(\bm{e}_{{\rm R},s}\times\bm{e}_{{\rm IR},s}\bigr)$ is a mode-dependent conversion vector that maps the mode-space angular momentum $\ell_Q=\hat{z}\cdot(\bm Q\times\dot{\bm Q})$ onto the physical sublattice-staggered PAM $\bm{L}_{\rm N}$. One can show that $\frac{Q_{\rm IR}}{Q_{\rm R}}$ is purely imaginary and
its sign is opposite for $\Psi_+$ and $\Psi_-$. The two modes thus have opposite mode-space angular momenta and opposite staggered angular momenta. 

Finally, reversing the N\'eel vector $\bm{N}$ reverses the sign of $\Omega$. This does not change eigenfrequencies as they depend on $\Omega^2$. The eigenvectors however are complex conjugated, reversing the phonon chirality and the staggered angular momentum. The AFCP chirality is thus locked to the N\'eel order:
\begin{equation}
    \bm N\to -\bm N
    \quad\Rightarrow\quad
    \Omega\to -\Omega
    \quad\Rightarrow\quad
    \bm{L}_{\rm N}\to -\bm{L}_{\rm N}.
\end{equation}
Reciprocally, the N\'eel vector can also be influenced by coherently excited phonons as demonstrated in the End Matter where the phonon and spin dynamics are solved simultaneously under external drive.

\begin{figure*}[t!]
\centering
\includegraphics[width= 17cm]{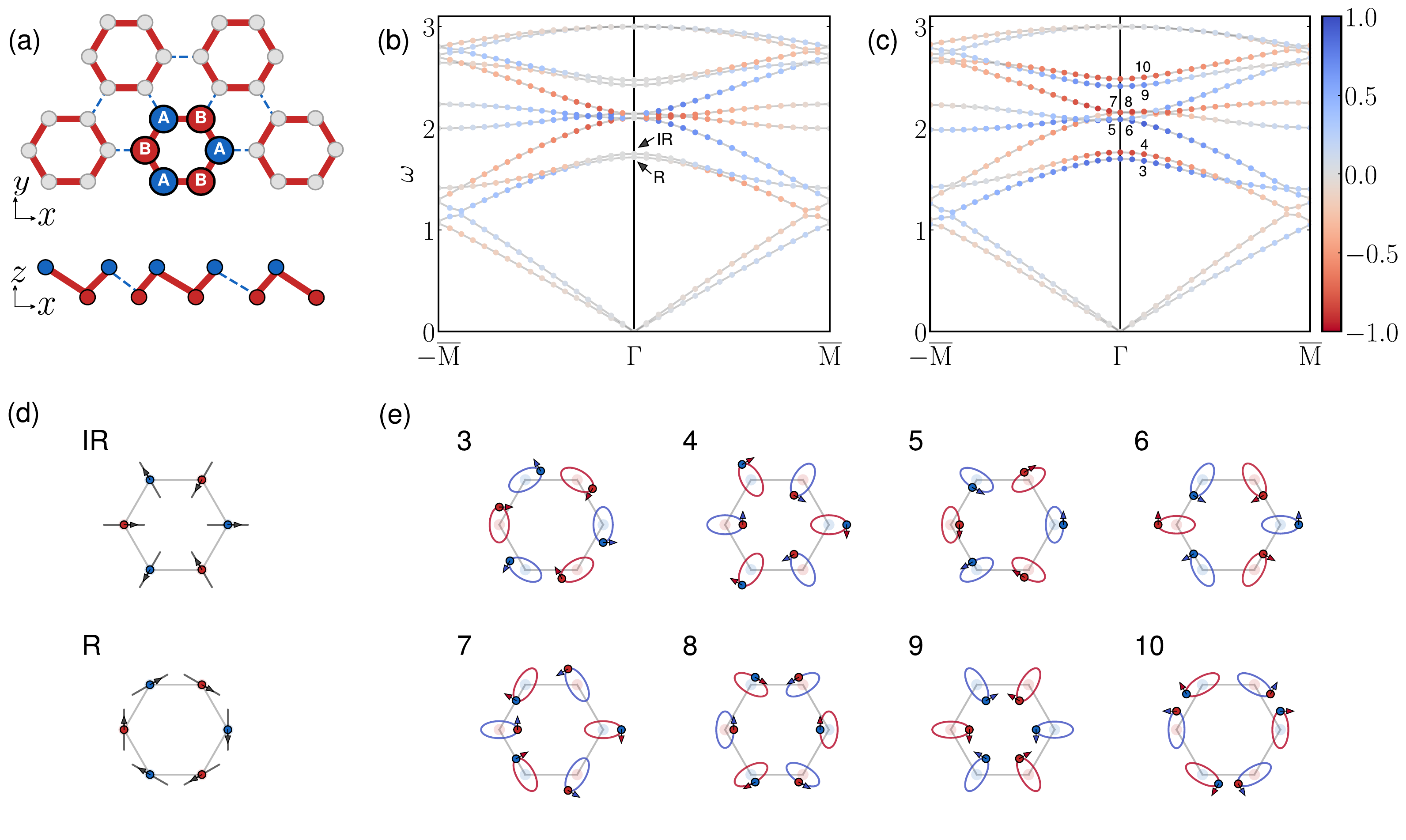}
\caption{(a) Distorted buckled hexagonal lattice formed by two atomic layers with six atoms in an unit cell. The atoms in each unit cell is grouped into A and B related by inversion operation. (b) Bare phonon band structure in the absence of the molecular Berry curvature. $\bm{L}_{\rm N}$ is encoded in color. Zone center optical modes are linearly polarized with $\bm{L}_{\rm N}=0$. Lattice vibration patterns of the lowest Raman (R) and IR modes are shown in panel (d).  (c) Phonon bands with molecular Berry curvature. (e) Lattice vibration patterns of eight AFCPs at $\Gamma$ point with nonzero $\bm{L}_{\rm N}$ numbered from $3$ to $10$ in panel (c). Frequency ($\omega$) is measured in units of $\sqrt{t_0/ma^2}$, where we have used the natural units, i.e., $t_0=1,m=1,\hbar=1$. }
\label{fig:bandstructure}
\end{figure*}

\emph{Molecular Berry curvature.---}
\label{MBC}
The symmetry analysis and minimal model are phenomenological and general. Microscopically, the Landau type coupling can be induced by the molecular Berry curvature, the effect of which has been studied in both gapped systems and semimetals \cite{Mead1979, Xiao2010, Qin2012,Saparov2022, Ren2024PRX, zhang2026general}. Within the adiabatic approximation, slow lattice motion drives the electronic ground state $|\Phi_0(\bm Q)\rangle$ along the phonon-coordinate manifold, generating a Berry phase correction to the phonon Lagrangian,
\begin{align}
    L_{\rm MBC} 
    = \sum_\mu A_\mu(\bm Q)\,\dot Q_\mu,
\end{align}
where the Berry connection is  $A_\mu = i\left\langle\Phi_0\middle|\partial_{Q_\mu} \Phi_0 \right\rangle$. 
Expanding the Berry connection near equilibrium with $\bm{Q}\sim 0$ gives $L_{\rm MBC} = \frac{1}{2}\sum_{\mu\nu} G_{\mu\nu} Q_\mu \dot Q_\nu$,
where the molecular Berry curvature tensor is defined as \cite{Saparov2022,Ren2025}
\begin{equation}\label{eq:MBC_formula}
    G_{\mu\nu}=\partial_{Q_\mu} A_{\nu} - \partial_{Q_\nu} A_{\mu} = 2\,\mathrm{Im}
    \left\langle \frac{\partial \Phi_0}{\partial Q_{\mu}}
    \,\bigg|\, \frac{\partial \Phi_0}{\partial Q_{\nu}} \right\rangle.
\end{equation}
For the Raman-IR normal-mode space, the antisymmetric $G_{\mu\nu}$ term reduces to the gyroscopic coupling, with 
\begin{align}
    \Omega = G_{Q_{\rm R}Q_{\rm IR}}.
\end{align}
Because $G_{\mu\nu}$ is odd under $\mathcal{T}$, it changes sign under N\'eel-vector reversal and is therefore locked to the N\'eel vector.

The Berry-phase correction to the lattice dynamics can be formally derived from the time-dependent variational principle by restricting the variational manifold to the instantaneous electronic ground state \cite{Saparov2022}. In the Cartesian coordinates with lattice displacements $\{\bm u\}$, the general Lagrangian reads
\begin{equation}\label{eq:Lagrangian}
    L = \sum_{s}\frac{M_s}{2} \dot{\bm{u}}_{s}^{2}
    + \bm A_{s}(\{\bm u\})\,\cdot \dot{\bm{u}}_{s} - V_{\rm eff}(\{\bm u\}),
\end{equation}
where $V_{\rm eff}(\{\bm u\})$ is the effective potential energy surface,$\bm A_{s} = i\left\langle\Phi_0\middle|\partial_{\bm u_{s}} \Phi_0 \right\rangle$ is the Berry connection in the atomic-coordinate space of the site $s$, and the corresponding Berry curvature is $G_{s\alpha,s'\alpha'} = \partial_{u_{s\alpha}} A_{s'\alpha'} - \partial_{u_{s'\alpha'}} A_{s\alpha}$. 
Varying this Lagrangian gives the equation of motion, which can be transformed to momentum space and solved for the phonon normal modes in the presence of the Berry-phase term~\cite{Saparov2022, DasSM}. 

\emph{A case study.---}\label{sec:model}
We now realize the AFCP mechanism in a concrete prototype lattice model. A $\mathcal{PT}$-symmetric lattice with two atoms per unit cell does not provide the IR-active partner mode required for the Raman-IR hybridization, so we consider instead a distorted honeycomb lattice as shown in Fig.~\ref{fig:bandstructure}(a), which carries both Raman and IR modes. The six-site unit cell can be partitioned into two inversion-related groups, $A$ and $B$, which occupy different out-of-plane heights forming two atomic layers. This partitioning will be used below to evaluate the sublattice-staggered angular momentum $\bm L_{\rm N}$.
We describe the bare lattice dynamics with a mass-spring model using bond-dependent stiffness matrices parameterized by longitudinal 
and transverse 
spring constants, rotated into the local bond frame. The stiffness matrices are different for intra- and inter-unit cell bonds.

In the absence of the molecular Berry curvature, the bare phonon modes are plotted in Fig.~\ref{fig:bandstructure}(b). 
Color encodes the $\bm L_{\rm N}$ of each mode. While the general phonon modes can carry a nonzero $\bm L_{\rm N}$, at the Brillouin zone center, the phonons are linearly polarized with $\bm L_{\rm N} = 0$ for the six Raman and four IR optical phonons. The vibration pattern of the lowest pair of optical modes are shown in Fig.~\ref{fig:bandstructure}(d), which are linearly polarized and carry no local angular momentum as shown in color. The inversion-even mode is identified as $Q_{\rm R}$ and the inversion-odd mode as $Q_{\rm IR}$. This pair provides a low-energy pair of opposite-parity optical modes that directly realize the bare Raman-IR doublet of the minimal model. 

The Berry-phase coupling emerges from the coupling between the phonons and the adiabatic evolution of the electronic wavefunction. We model the electronic structure with a tight-binding Hamiltonian on the same lattice, with electronic hoppings, a staggered exchange field from the AFM order, and spin-orbit couplings that break the spin-rotational symmetry and is crucial for generating nonzero molecular Berry curvature. 
The tight-binding Hamiltonian reads~\cite{liu2011low, mccann2013electronic, DasSM}
\begin{align}\label{eq:elec_hamiltonian}
    \mathcal{H}_{el} =& \left[ -\sum_{\langle i,j \rangle, \sigma} t_0 (1 + \delta_{ij}) c_{i\sigma}^{\dagger} c_{j\sigma} \right] + \sum_{i, \alpha, \beta} N_i c_{i\alpha}^{\dagger} (\sigma_z)_{\alpha\beta} c_{i\beta} \notag \\
    &- i \sum_{\langle\langle i,j \rangle\rangle, \alpha, \beta} t_{\rm IR}  \mu_{ij} \left[ c_{i\alpha}^{\dagger} \left( \boldsymbol{\sigma} \times \hat{\bm{d}}_{ij} \right)_{z}^{\alpha\beta} c_{j\beta}\right] , \notag
\end{align}
where $\langle i,j \rangle$ runs over three nearest neighbor sites and $\langle\langle i,j \rangle\rangle$ involves six next-nearest neighbors. $\delta_{ij}=\pm \delta$ represents intra- and inter-unit cell hoppings depending on the distortion. $N_i=\pm N$ on $A/B$ sites is the staggered exchange field, $\mu_{ij}=\pm 1$ for A/B sublattices, and $t_{\rm IR}$ is the intrinsic Rashba spin-orbit coupling~\cite{liu2011low}.
The hopping-integral amplitudes depend on the bond lengths as \(t_0=t_{0}^{0}\exp[-\beta_0(r/r_{0}-1)]\) where $r$ is the bond length with lattice displacement $\bm u$, $r_{0}$ is the equilibrium bond length, and $\beta_0=3$ is the electron-phonon coupling parameter~\cite{mccann2013electronic}. The unit vector $\hat{\bm{d}}_{ij}$ points from site $j$ to $i$, which also depends on $\bm u$ and contributes to the electron-phonon coupling \cite{DasSM}.
The electronic spectrum is gapped \cite{DasSM}, and the ground state $|\Phi_0\rangle$ is a Slater determinant of occupied single-particle states, while $|\Phi_n\rangle$ denotes the corresponding excited Slater determinants.
Inserting a complete set of excited states gives
\begin{equation}\label{eq:MBC}
    G_{s\alpha,s'\alpha'} = -2\,\mathrm{Im}\sum_{n\neq 0}
    \frac{\langle\Phi_0|\partial_{u_{s\alpha}}\mathcal{H}_{el}|\Phi_n\rangle
          \langle\Phi_n|\partial_{u_{s'\alpha'}}\mathcal{H}_{el}|\Phi_0\rangle}
         {(E_n - E_0)^2}, \notag
\end{equation}
where $E_n$ are the excited-state energies. 

\begin{table}[t]
\begin{ruledtabular}
\begin{tabular}{cccccc}
Mode & $\omega$ & $\it{l}_A$ & $\it{l}_B$ & $\bm{L}_{\rm N}$ \\
\colrule
3  & 1.70 &  0.420 & $-$0.420  &  0.840 \\
4  & 1.76 & $-$0.415 &  0.415  & $-$0.831 \\
5  & 2.09 &  0.391 &  $-$0.391  &  0.782 \\
6  & 2.09 & 0.391 & $-$0.391  &  0.782 \\
7  & 2.16 &  $-$0.386 &  0.386  & $-$0.772 \\
8  & 2.16 & $-$0.386 & 0.386  & $-$0.772 \\
9  & 2.41 &  0.368 & $-$0.368  &  0.737 \\
10 & 2.49 & $-$0.363 &  0.363  & $-$0.727 \\
\end{tabular}
\end{ruledtabular}
\caption{\label{tab:angular_momentum} Lattice-resolved angular momentum of the $\Gamma$ point eigenmodes. The table details the mode index, frequency ($\omega$), sublattice-resolved PAM ($\it{l}_A$, $\it{l}_B$), and the staggered PAM ($\bm{L}_{\rm N}$) in units of $\hbar$.}
\end{table}

We then solve the Berry-phase-corrected lattice dynamics to obtain the phonon dispersion as shown in Fig.~\ref{fig:bandstructure}(c). The resulting modes retain zero total angular momentum, while developing a finite sublattice-staggered angular momentum $\bm L_{\rm N}$, encoded in color. The zone-center modes are no longer linearly polarized: the lattice sites execute local elliptical motions whose angular momenta alternate between the inversion-related groups $A$ and $B$ as plotted in Fig.~\ref{fig:bandstructure}(e). 

\emph{Degenerate AFCPs.---} Figure~\ref{fig:bandstructure}(c) also present degenerate AFCPs arising from the coupling between the doubly degenerate Raman and IR sectors. The minimal description thus extends naturally to a four-dimensional parity-polarization subspace. Writing $\bm Q_{\rm R}=(Q_{\rm Rx},Q_{\rm Ry})^{\rm T}$ and $\bm Q_{\rm IR}=(Q_{\rm IRx},Q_{\rm IRy})^{\rm T}$, we define
\begin{equation}
    \bm Q=(Q_{\rm Rx},Q_{\rm Ry},Q_{\rm IRx},Q_{\rm IRy})^{\rm T}.
\end{equation}
The bare scalar-potential dynamics is governed by
\begin{equation}
    D_0=(\omega_0^2\sigma_0+\Delta\sigma_z)\otimes\tau_0,
\end{equation}
where $\sigma_i$ act in the Raman-IR parity space and $\tau_i$ in the in-plane polarization space. Within each parity sector, the $x$ and $y$ components can be recombined into chiral phonons carrying opposite total angular momenta, but vanishing sublattice-staggered angular momentum.

The $\mathcal{PT}$-allowed molecular Berry curvature correction is again gyroscopic, but now acts in the extended space. As shown in the Supplemental Materials \cite{DasSM}, $\mathcal{PT}$ symmetry forbids intra-parity gyroscopic terms, while the threefold rotation constrains the inter-parity coupling to the canonical form used below,
\begin{equation}\label{eq:4x4eff}
    L_{\rm eff}^{(4)}
    = \frac{1}{2}\dot{\bm Q}^{\,2}
    - \frac{1}{2}\bm Q^{\rm T}D_0\bm Q
    + \frac{\Omega}{2}\bm Q^{\rm T}(\sigma_x\otimes\varepsilon)\dot{\bm Q},
\end{equation}
with $\varepsilon=\begin{pmatrix}0&1\\-1&0\end{pmatrix}$. By passing to the circular basis,
\begin{equation}
    Q_{{\rm R}\pm}=\frac{Q_{\rm Rx}\pm iQ_{\rm Ry}}{\sqrt{2}},\qquad
    Q_{{\rm IR}\pm}=\frac{Q_{\rm IRx}\pm iQ_{\rm IRy}}{\sqrt{2}},
\end{equation}
we block diagonalize the mode equation into two decoupled blocks, $(Q_{{\rm R}+},Q_{{\rm IR}+})$ and $(Q_{{\rm R}-},Q_{{\rm IR}-})$. Each block is governed by the same $2\times2$ gyroscopic coupling as in the nondegenerate minimal model. 

The two circular-polarization blocks are related to each other by the $\mathcal{PT}$ symmetry. Thus, their eigenfrequencies are identical, producing two doubly-degenerate phonon branches. Within each branch, the two degenerate modes are $\mathcal{PT}$ partners contributed from the two blocks, separately. They have the same sublattice-staggered angular momentum $\bm L_{\rm N}$, but opposite net angular momentum $\bm L$. 
Within a degenerate subspace, the eigenvectors are not unique. We find that a generic superposition of these two degenerate modes changes the net angular momentum $\bm L$ while preserving the sublattice-staggered angular momentum $\bm L_{\rm N}$. We thus define $\bm L_{\rm N} = \frac{1}{2}{\rm Tr}\hat{P}\hat{\bm L}_{\rm N}\hat{P}$ where $\hat{P}$ is the projection operator onto the degenerate subspace with mass-weighted polarization vectors. $\hat{\bm L}_{\rm N}= \bigoplus_s \eta_s \hat{\bm J}_{s}$ is the sublattice-staggered angular momentum operator with $\eta_s = \pm 1$ for two sets of sublattices and $\hat{\bm J}_{s}$ the angular momentum operator of atom $s$.
$\mathcal{PT}$-invariant superpositions are special, which shows zero total angular momentum and can be selected by weak symmetry
breaking, such as a small lattice distortion from strain that breaks the three-fold rotational symmetry in our model.



\begin{figure}[t!]
\centering
\includegraphics[width=8.5cm]{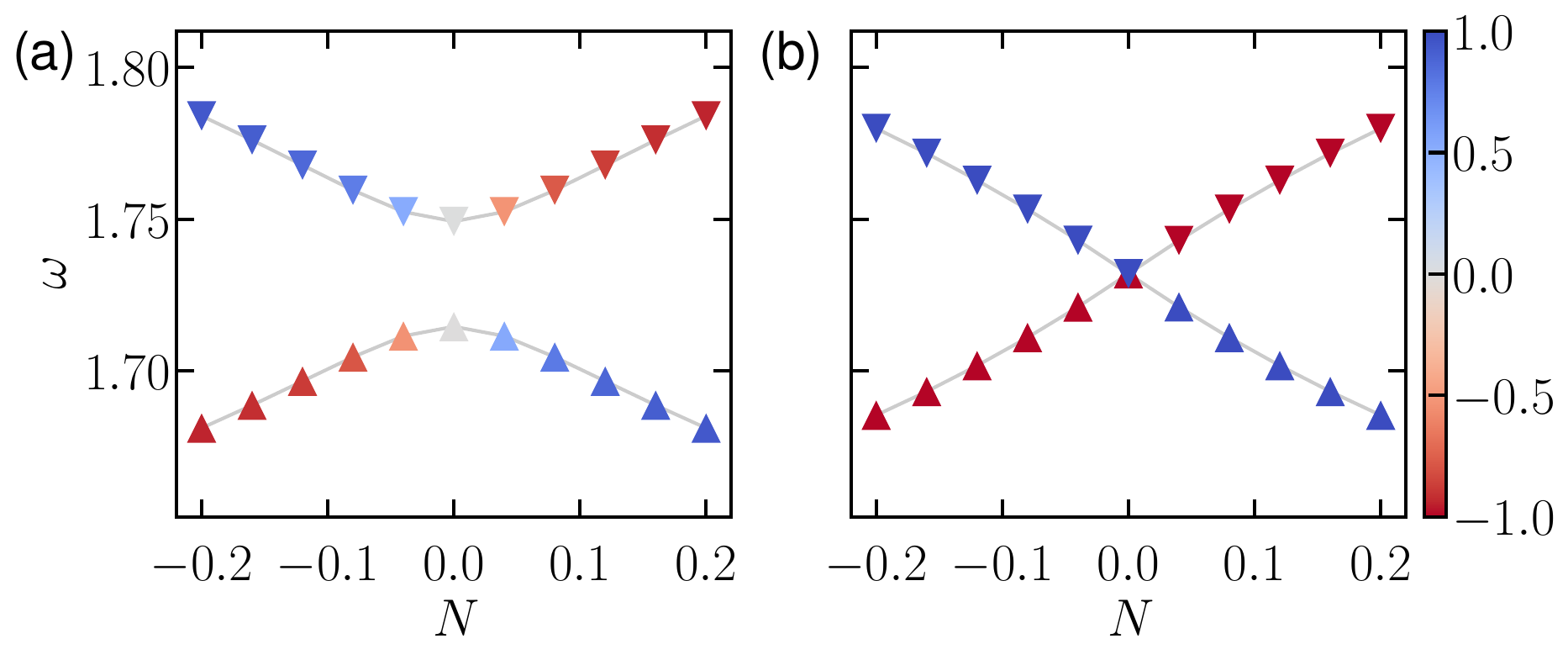}
\caption{Lowest-energy Raman and IR modes vs AFM exchange field $N$. The Raman and IR modes are nondegenerate in (a) whereas are set to be accidentally degenerate in (b) at $N=0$. Color encodes $\bm L_{\rm N}$ in units of $\hbar$. Both panels are evaluated in the presence of molecular Berry curvature. We set the electronic Hamiltonian parameters to $t_0=1, t_{\rm IR}=0.2$.}
\label{fig:splitting}
\end{figure}

\emph{N\'eel-vector locking and chirality switching.---}
Reversing the N\'eel vector reverses the AFCP chirality. Figure~\ref{fig:splitting}(a) shows the evolution of the lowest two AFCP branches as a function of the AFM exchange field $N$, with the color scale encoding the sublattice-staggered PAM $\bm L_{\rm N}$. $\bm L_{\rm N}$ shows sensitive dependence on the ratio between the molecular Berry curvature induced coupling and the Raman-IR detuning. At $N=0$, the electronic system is time-reversal symmetric, the molecular Berry curvature vanishes, and the two zone-center modes are linearly polarized Raman and IR phonons with $\bm L_{\rm N}=0$. Turning on $N$ activates the Berry curvature, hybridizes the Raman and IR modes, and generates AFCPs with finite $\bm L_{\rm N}$ that increases from zero and can reach the order of magnitude of $\hbar$ when the molecular Berry curvature-induced coupling is comparable to the detuning. For opposite signs of $N$, the phonon frequencies remain unchanged, while the eigenvectors are complex conjugated and therefore carry opposite $\bm L_{\rm N}$.

To isolate the molecular Berry curvature-induced splitting, we also consider the case where the bare Raman and IR modes are accidentally degenerate as shown in Fig.~\ref{fig:splitting}(b). The resulting splitting of the two AFCP branches directly measures the gyroscopic coupling generated by the molecular Berry curvature. The results indicate that the gyroscopic coupling is odd in $N$, showing a nearly linear dependence in this situation.

\emph{Summary and discussion.---}
\label{sec:conc}
We have shown that the sublattice-staggered PAM generates a conjugate field to the N\'eel vector, showing a Landau-type coupling.
Microscopically, the N\'eel-vector-locked molecular Berry curvature can induce such a coupling, which generates a gyroscopic hybridization between opposite-parity Raman and IR phonons. This coupling converts linearly polarized bare modes into AFCPs with staggered PAM $\bm L_{\rm N}$, and the sign of $\bm L_{\rm N}$ reverses as the N\'eel vector is reversed, providing a phononic quantity directly tied to the antiferromagnetic order.
The conjugate field to the N\'eel vector enables a phononic route to access compensated antiferromagnetic order through lattice dynamics as illustrated in the End Matter \cite{Wu2022,Zhang2025spinphonon}. 

The effects of the Landau-type coupling is controlled by the Raman-IR phonon-frequency detuning and the molecular-Berry-curvature amplitude. Materials with near-resonant opposite-parity phonons are therefore natural candidates \cite{Kargar2020,Cui2023,Dixit2026}. The molecular Berry curvature can be enhanced by strong spin-orbit coupling and low-lying excited states, such as the topological antiferromagnets~\cite{otrokov2019prediction, tang2016dirac} and magnets with unquenched orbitals~\cite{khusyainov2023ultrafast, lee2023giant, dhakal2024hybrid}.

\emph{Acknowledgments.---} S.K.D would like to acknowledge fruitful discussions with Chao-Xing Liu and Qian Niu. This research was supported by the US National Science Foundation (NSF) Grant No. 2531815.

\bibliography{references}

\begin{thebibliography}{85}%
\makeatletter
\providecommand \@ifxundefined [1]{%
 \@ifx{#1\undefined}
}%
\providecommand \@ifnum [1]{%
 \ifnum #1\expandafter \@firstoftwo
 \else \expandafter \@secondoftwo
 \fi
}%
\providecommand \@ifx [1]{%
 \ifx #1\expandafter \@firstoftwo
 \else \expandafter \@secondoftwo
 \fi
}%
\providecommand \natexlab [1]{#1}%
\providecommand \enquote  [1]{``#1''}%
\providecommand \bibnamefont  [1]{#1}%
\providecommand \bibfnamefont [1]{#1}%
\providecommand \citenamefont [1]{#1}%
\providecommand \href@noop [0]{\@secondoftwo}%
\providecommand \href [0]{\begingroup \@sanitize@url \@href}%
\providecommand \@href[1]{\@@startlink{#1}\@@href}%
\providecommand \@@href[1]{\endgroup#1\@@endlink}%
\providecommand \@sanitize@url [0]{\catcode `\\12\catcode `\$12\catcode
  `\&12\catcode `\#12\catcode `\^12\catcode `\_12\catcode `\%12\relax}%
\providecommand \@@startlink[1]{}%
\providecommand \@@endlink[0]{}%
\providecommand \url  [0]{\begingroup\@sanitize@url \@url }%
\providecommand \@url [1]{\endgroup\@href {#1}{\urlprefix }}%
\providecommand \urlprefix  [0]{URL }%
\providecommand \Eprint [0]{\href }%
\providecommand \doibase [0]{https://doi.org/}%
\providecommand \selectlanguage [0]{\@gobble}%
\providecommand \bibinfo  [0]{\@secondoftwo}%
\providecommand \bibfield  [0]{\@secondoftwo}%
\providecommand \translation [1]{[#1]}%
\providecommand \BibitemOpen [0]{}%
\providecommand \bibitemStop [0]{}%
\providecommand \bibitemNoStop [0]{.\EOS\space}%
\providecommand \EOS [0]{\spacefactor3000\relax}%
\providecommand \BibitemShut  [1]{\csname bibitem#1\endcsname}%
\let\auto@bib@innerbib\@empty
\bibitem [{\citenamefont {Gomonay}\ and\ \citenamefont
  {Loktev}(2014)}]{Gomonay2014}%
  \BibitemOpen
  \bibfield  {author} {\bibinfo {author} {\bibfnamefont {E.~V.}\ \bibnamefont
  {Gomonay}}\ and\ \bibinfo {author} {\bibfnamefont {V.~M.}\ \bibnamefont
  {Loktev}},\ }\bibfield  {title} {\bibinfo {title} {Spintronics of
  antiferromagnetic systems (review article)},\ }\href
  {https://doi.org/10.1063/1.4862467} {\bibfield  {journal} {\bibinfo
  {journal} {Low Temp. Phys.}\ }\textbf {\bibinfo {volume} {40}},\ \bibinfo
  {pages} {17–35} (\bibinfo {year} {2014})}\BibitemShut {NoStop}%
\bibitem [{\citenamefont {Baltz}\ \emph {et~al.}(2018)\citenamefont {Baltz},
  \citenamefont {Manchon}, \citenamefont {Tsoi}, \citenamefont {Moriyama},
  \citenamefont {Ono},\ and\ \citenamefont {Tserkovnyak}}]{Baltz2018}%
  \BibitemOpen
  \bibfield  {author} {\bibinfo {author} {\bibfnamefont {V.}~\bibnamefont
  {Baltz}}, \bibinfo {author} {\bibfnamefont {A.}~\bibnamefont {Manchon}},
  \bibinfo {author} {\bibfnamefont {M.}~\bibnamefont {Tsoi}}, \bibinfo {author}
  {\bibfnamefont {T.}~\bibnamefont {Moriyama}}, \bibinfo {author}
  {\bibfnamefont {T.}~\bibnamefont {Ono}},\ and\ \bibinfo {author}
  {\bibfnamefont {Y.}~\bibnamefont {Tserkovnyak}},\ }\bibfield  {title}
  {\bibinfo {title} {Antiferromagnetic spintronics},\ }\href
  {https://doi.org/10.1103/RevModPhys.90.015005} {\bibfield  {journal}
  {\bibinfo  {journal} {Rev. Mod. Phys.}\ }\textbf {\bibinfo {volume} {90}},\
  \bibinfo {pages} {015005} (\bibinfo {year} {2018})}\BibitemShut {NoStop}%
\bibitem [{\citenamefont {Železný}\ \emph {et~al.}(2018)\citenamefont
  {Železný}, \citenamefont {Wadley}, \citenamefont {Olejník}, \citenamefont
  {Hoffmann},\ and\ \citenamefont {Ohno}}]{elezn2018}%
  \BibitemOpen
  \bibfield  {author} {\bibinfo {author} {\bibfnamefont {J.}~\bibnamefont
  {Železný}}, \bibinfo {author} {\bibfnamefont {P.}~\bibnamefont {Wadley}},
  \bibinfo {author} {\bibfnamefont {K.}~\bibnamefont {Olejník}}, \bibinfo
  {author} {\bibfnamefont {A.}~\bibnamefont {Hoffmann}},\ and\ \bibinfo
  {author} {\bibfnamefont {H.}~\bibnamefont {Ohno}},\ }\bibfield  {title}
  {\bibinfo {title} {Spin transport and spin torque in antiferromagnetic
  devices},\ }\href {https://doi.org/10.1038/s41567-018-0062-7} {\bibfield
  {journal} {\bibinfo  {journal} {Nat. Phys.}\ }\textbf {\bibinfo {volume}
  {14}},\ \bibinfo {pages} {220–228} (\bibinfo {year} {2018})}\BibitemShut
  {NoStop}%
\bibitem [{\citenamefont {Manchon}\ \emph {et~al.}(2019)\citenamefont
  {Manchon}, \citenamefont {\ifmmode~\check{Z}\else \v{Z}\fi{}elezn\'y},
  \citenamefont {Miron}, \citenamefont {Jungwirth}, \citenamefont {Sinova},
  \citenamefont {Thiaville}, \citenamefont {Garello},\ and\ \citenamefont
  {Gambardella}}]{Manchon19}%
  \BibitemOpen
  \bibfield  {author} {\bibinfo {author} {\bibfnamefont {A.}~\bibnamefont
  {Manchon}}, \bibinfo {author} {\bibfnamefont {J.}~\bibnamefont
  {\ifmmode~\check{Z}\else \v{Z}\fi{}elezn\'y}}, \bibinfo {author}
  {\bibfnamefont {I.~M.}\ \bibnamefont {Miron}}, \bibinfo {author}
  {\bibfnamefont {T.}~\bibnamefont {Jungwirth}}, \bibinfo {author}
  {\bibfnamefont {J.}~\bibnamefont {Sinova}}, \bibinfo {author} {\bibfnamefont
  {A.}~\bibnamefont {Thiaville}}, \bibinfo {author} {\bibfnamefont
  {K.}~\bibnamefont {Garello}},\ and\ \bibinfo {author} {\bibfnamefont
  {P.}~\bibnamefont {Gambardella}},\ }\bibfield  {title} {\bibinfo {title}
  {Current-induced spin-orbit torques in ferromagnetic and antiferromagnetic
  systems},\ }\href {https://doi.org/10.1103/RevModPhys.91.035004} {\bibfield
  {journal} {\bibinfo  {journal} {Rev. Mod. Phys.}\ }\textbf {\bibinfo {volume}
  {91}},\ \bibinfo {pages} {035004} (\bibinfo {year} {2019})}\BibitemShut
  {NoStop}%
\bibitem [{\citenamefont {Han}\ \emph {et~al.}(2023)\citenamefont {Han},
  \citenamefont {Cheng}, \citenamefont {Liu}, \citenamefont {Ohno},\ and\
  \citenamefont {Fukami}}]{Han2023}%
  \BibitemOpen
  \bibfield  {author} {\bibinfo {author} {\bibfnamefont {J.}~\bibnamefont
  {Han}}, \bibinfo {author} {\bibfnamefont {R.}~\bibnamefont {Cheng}}, \bibinfo
  {author} {\bibfnamefont {L.}~\bibnamefont {Liu}}, \bibinfo {author}
  {\bibfnamefont {H.}~\bibnamefont {Ohno}},\ and\ \bibinfo {author}
  {\bibfnamefont {S.}~\bibnamefont {Fukami}},\ }\bibfield  {title} {\bibinfo
  {title} {Coherent antiferromagnetic spintronics},\ }\href
  {https://doi.org/10.1038/s41563-023-01492-6} {\bibfield  {journal} {\bibinfo
  {journal} {Nat. Mater.}\ }\textbf {\bibinfo {volume} {22}},\ \bibinfo {pages}
  {684–695} (\bibinfo {year} {2023})}\BibitemShut {NoStop}%
\bibitem [{\citenamefont {René}\ \emph {et~al.}(2025)\citenamefont {René},
  \citenamefont {Levchuk}, \citenamefont {Abdelsamie}, \citenamefont {Li},
  \citenamefont {Dufour}, \citenamefont {Chaudron}, \citenamefont {Godel},
  \citenamefont {Gemeiner}, \citenamefont {Dkhil}, \citenamefont {Moussy},
  \citenamefont {Bouzehouane}, \citenamefont {Fusil}, \citenamefont {Garcia},
  \citenamefont {Viret},\ and\ \citenamefont {Chauleau}}]{Rene2025}%
  \BibitemOpen
  \bibfield  {author} {\bibinfo {author} {\bibfnamefont {S.}~\bibnamefont
  {René}}, \bibinfo {author} {\bibfnamefont {A.}~\bibnamefont {Levchuk}},
  \bibinfo {author} {\bibfnamefont {A.}~\bibnamefont {Abdelsamie}}, \bibinfo
  {author} {\bibfnamefont {Z.}~\bibnamefont {Li}}, \bibinfo {author}
  {\bibfnamefont {P.}~\bibnamefont {Dufour}}, \bibinfo {author} {\bibfnamefont
  {A.}~\bibnamefont {Chaudron}}, \bibinfo {author} {\bibfnamefont
  {F.}~\bibnamefont {Godel}}, \bibinfo {author} {\bibfnamefont
  {P.}~\bibnamefont {Gemeiner}}, \bibinfo {author} {\bibfnamefont
  {B.}~\bibnamefont {Dkhil}}, \bibinfo {author} {\bibfnamefont {J.-B.}\
  \bibnamefont {Moussy}}, \bibinfo {author} {\bibfnamefont {K.}~\bibnamefont
  {Bouzehouane}}, \bibinfo {author} {\bibfnamefont {S.}~\bibnamefont {Fusil}},
  \bibinfo {author} {\bibfnamefont {V.}~\bibnamefont {Garcia}}, \bibinfo
  {author} {\bibfnamefont {M.}~\bibnamefont {Viret}},\ and\ \bibinfo {author}
  {\bibfnamefont {J.-Y.}\ \bibnamefont {Chauleau}},\ }\bibfield  {title}
  {\bibinfo {title} {Terahertz antiferromagnetic dynamics induced by ultrafast
  spin currents},\ }\href {https://doi.org/10.1126/sciadv.adx1107} {\bibfield
  {journal} {\bibinfo  {journal} {Sci. Adv.}\ }\textbf {\bibinfo {volume}
  {11}},\ \bibinfo {pages} {1107} (\bibinfo {year} {2025})}\BibitemShut
  {NoStop}%
\bibitem [{\citenamefont {Wadley}\ \emph {et~al.}(2016)\citenamefont {Wadley},
  \citenamefont {Howells}, \citenamefont {Železný}, \citenamefont {Andrews},
  \citenamefont {Hills}, \citenamefont {Campion}, \citenamefont {Novák},
  \citenamefont {Olejník}, \citenamefont {Maccherozzi}, \citenamefont {Dhesi},
  \citenamefont {Martin}, \citenamefont {Wagner}, \citenamefont {Wunderlich},
  \citenamefont {Freimuth}, \citenamefont {Mokrousov}, \citenamefont {Kuneš},
  \citenamefont {Chauhan}, \citenamefont {Grzybowski}, \citenamefont
  {Rushforth}, \citenamefont {Edmonds}, \citenamefont {Gallagher},\ and\
  \citenamefont {Jungwirth}}]{Wadley2016}%
  \BibitemOpen
  \bibfield  {author} {\bibinfo {author} {\bibfnamefont {P.}~\bibnamefont
  {Wadley}}, \bibinfo {author} {\bibfnamefont {B.}~\bibnamefont {Howells}},
  \bibinfo {author} {\bibfnamefont {J.}~\bibnamefont {Železný}}, \bibinfo
  {author} {\bibfnamefont {C.}~\bibnamefont {Andrews}}, \bibinfo {author}
  {\bibfnamefont {V.}~\bibnamefont {Hills}}, \bibinfo {author} {\bibfnamefont
  {R.~P.}\ \bibnamefont {Campion}}, \bibinfo {author} {\bibfnamefont
  {V.}~\bibnamefont {Novák}}, \bibinfo {author} {\bibfnamefont
  {K.}~\bibnamefont {Olejník}}, \bibinfo {author} {\bibfnamefont
  {F.}~\bibnamefont {Maccherozzi}}, \bibinfo {author} {\bibfnamefont {S.~S.}\
  \bibnamefont {Dhesi}}, \bibinfo {author} {\bibfnamefont {S.~Y.}\ \bibnamefont
  {Martin}}, \bibinfo {author} {\bibfnamefont {T.}~\bibnamefont {Wagner}},
  \bibinfo {author} {\bibfnamefont {J.}~\bibnamefont {Wunderlich}}, \bibinfo
  {author} {\bibfnamefont {F.}~\bibnamefont {Freimuth}}, \bibinfo {author}
  {\bibfnamefont {Y.}~\bibnamefont {Mokrousov}}, \bibinfo {author}
  {\bibfnamefont {J.}~\bibnamefont {Kuneš}}, \bibinfo {author} {\bibfnamefont
  {J.~S.}\ \bibnamefont {Chauhan}}, \bibinfo {author} {\bibfnamefont {M.~J.}\
  \bibnamefont {Grzybowski}}, \bibinfo {author} {\bibfnamefont {A.~W.}\
  \bibnamefont {Rushforth}}, \bibinfo {author} {\bibfnamefont {K.~W.}\
  \bibnamefont {Edmonds}}, \bibinfo {author} {\bibfnamefont {B.~L.}\
  \bibnamefont {Gallagher}},\ and\ \bibinfo {author} {\bibfnamefont
  {T.}~\bibnamefont {Jungwirth}},\ }\bibfield  {title} {\bibinfo {title}
  {Electrical switching of an antiferromagnet},\ }\href
  {https://doi.org/10.1126/science.aab1031} {\bibfield  {journal} {\bibinfo
  {journal} {Science}\ }\textbf {\bibinfo {volume} {351}},\ \bibinfo {pages}
  {587–590} (\bibinfo {year} {2016})}\BibitemShut {NoStop}%
\bibitem [{\citenamefont {Moriyama}\ \emph {et~al.}(2018)\citenamefont
  {Moriyama}, \citenamefont {Oda}, \citenamefont {Ohkochi}, \citenamefont
  {Kimata},\ and\ \citenamefont {Ono}}]{Moriyama2018}%
  \BibitemOpen
  \bibfield  {author} {\bibinfo {author} {\bibfnamefont {T.}~\bibnamefont
  {Moriyama}}, \bibinfo {author} {\bibfnamefont {K.}~\bibnamefont {Oda}},
  \bibinfo {author} {\bibfnamefont {T.}~\bibnamefont {Ohkochi}}, \bibinfo
  {author} {\bibfnamefont {M.}~\bibnamefont {Kimata}},\ and\ \bibinfo {author}
  {\bibfnamefont {T.}~\bibnamefont {Ono}},\ }\bibfield  {title} {\bibinfo
  {title} {Spin torque control of antiferromagnetic moments in {NiO}},\ }\href
  {https://doi.org/10.1038/s41598-018-32508-w} {\bibfield  {journal} {\bibinfo
  {journal} {Sci. Rep.}\ }\textbf {\bibinfo {volume} {8}},\ \bibinfo {pages}
  {14167} (\bibinfo {year} {2018})}\BibitemShut {NoStop}%
\bibitem [{\citenamefont {Chiang}\ \emph {et~al.}(2019)\citenamefont {Chiang},
  \citenamefont {Huang}, \citenamefont {Qu}, \citenamefont {Wu},\ and\
  \citenamefont {Chien}}]{Chiang2019}%
  \BibitemOpen
  \bibfield  {author} {\bibinfo {author} {\bibfnamefont {C.~C.}\ \bibnamefont
  {Chiang}}, \bibinfo {author} {\bibfnamefont {S.~Y.}\ \bibnamefont {Huang}},
  \bibinfo {author} {\bibfnamefont {D.}~\bibnamefont {Qu}}, \bibinfo {author}
  {\bibfnamefont {P.~H.}\ \bibnamefont {Wu}},\ and\ \bibinfo {author}
  {\bibfnamefont {C.~L.}\ \bibnamefont {Chien}},\ }\bibfield  {title} {\bibinfo
  {title} {{Absence of Evidence of Electrical Switching of the
  Antiferromagnetic N\'eel Vector}},\ }\href
  {https://doi.org/10.1103/PhysRevLett.123.227203} {\bibfield  {journal}
  {\bibinfo  {journal} {Phys. Rev. Lett.}\ }\textbf {\bibinfo {volume} {123}},\
  \bibinfo {pages} {227203} (\bibinfo {year} {2019})}\BibitemShut {NoStop}%
\bibitem [{\citenamefont {Strohm}\ \emph {et~al.}(2005)\citenamefont {Strohm},
  \citenamefont {Rikken},\ and\ \citenamefont {Wyder}}]{Strohm2005}%
  \BibitemOpen
  \bibfield  {author} {\bibinfo {author} {\bibfnamefont {C.}~\bibnamefont
  {Strohm}}, \bibinfo {author} {\bibfnamefont {G.~L. J.~A.}\ \bibnamefont
  {Rikken}},\ and\ \bibinfo {author} {\bibfnamefont {P.}~\bibnamefont
  {Wyder}},\ }\bibfield  {title} {\bibinfo {title} {{Phenomenological Evidence
  for the Phonon Hall Effect}},\ }\href
  {https://doi.org/10.1103/PhysRevLett.95.155901} {\bibfield  {journal}
  {\bibinfo  {journal} {Phys. Rev. Lett.}\ }\textbf {\bibinfo {volume} {95}},\
  \bibinfo {pages} {155901} (\bibinfo {year} {2005})}\BibitemShut {NoStop}%
\bibitem [{\citenamefont {Zhang}\ \emph {et~al.}(2010)\citenamefont {Zhang},
  \citenamefont {Ren}, \citenamefont {Wang},\ and\ \citenamefont
  {Li}}]{Zhang2010}%
  \BibitemOpen
  \bibfield  {author} {\bibinfo {author} {\bibfnamefont {L.}~\bibnamefont
  {Zhang}}, \bibinfo {author} {\bibfnamefont {J.}~\bibnamefont {Ren}}, \bibinfo
  {author} {\bibfnamefont {J.-S.}\ \bibnamefont {Wang}},\ and\ \bibinfo
  {author} {\bibfnamefont {B.}~\bibnamefont {Li}},\ }\bibfield  {title}
  {\bibinfo {title} {{Topological Nature of the Phonon Hall Effect}},\ }\href
  {https://doi.org/10.1103/PhysRevLett.105.225901} {\bibfield  {journal}
  {\bibinfo  {journal} {Phys. Rev. Lett.}\ }\textbf {\bibinfo {volume} {105}},\
  \bibinfo {pages} {225901} (\bibinfo {year} {2010})}\BibitemShut {NoStop}%
\bibitem [{\citenamefont {Qin}\ \emph {et~al.}(2012)\citenamefont {Qin},
  \citenamefont {Zhou},\ and\ \citenamefont {Shi}}]{Qin2012}%
  \BibitemOpen
  \bibfield  {author} {\bibinfo {author} {\bibfnamefont {T.}~\bibnamefont
  {Qin}}, \bibinfo {author} {\bibfnamefont {J.}~\bibnamefont {Zhou}},\ and\
  \bibinfo {author} {\bibfnamefont {J.}~\bibnamefont {Shi}},\ }\bibfield
  {title} {\bibinfo {title} {{Berry curvature and the phonon Hall effect}},\
  }\href {https://doi.org/10.1103/PhysRevB.86.104305} {\bibfield  {journal}
  {\bibinfo  {journal} {Phys. Rev. B}\ }\textbf {\bibinfo {volume} {86}},\
  \bibinfo {pages} {104305} (\bibinfo {year} {2012})}\BibitemShut {NoStop}%
\bibitem [{\citenamefont {Zhang}\ and\ \citenamefont {Niu}(2014)}]{Zhang2014}%
  \BibitemOpen
  \bibfield  {author} {\bibinfo {author} {\bibfnamefont {L.}~\bibnamefont
  {Zhang}}\ and\ \bibinfo {author} {\bibfnamefont {Q.}~\bibnamefont {Niu}},\
  }\bibfield  {title} {\bibinfo {title} {{Angular Momentum of Phonons and the
  Einstein--de Haas Effect}},\ }\href
  {https://doi.org/10.1103/PhysRevLett.112.085503} {\bibfield  {journal}
  {\bibinfo  {journal} {Phys. Rev. Lett.}\ }\textbf {\bibinfo {volume} {112}},\
  \bibinfo {pages} {085503} (\bibinfo {year} {2014})}\BibitemShut {NoStop}%
\bibitem [{\citenamefont {Chen}\ \emph {et~al.}(2015)\citenamefont {Chen},
  \citenamefont {Zheng}, \citenamefont {Fuhrer},\ and\ \citenamefont
  {Yan}}]{Chen2015}%
  \BibitemOpen
  \bibfield  {author} {\bibinfo {author} {\bibfnamefont {S.-Y.}\ \bibnamefont
  {Chen}}, \bibinfo {author} {\bibfnamefont {C.}~\bibnamefont {Zheng}},
  \bibinfo {author} {\bibfnamefont {M.~S.}\ \bibnamefont {Fuhrer}},\ and\
  \bibinfo {author} {\bibfnamefont {J.}~\bibnamefont {Yan}},\ }\bibfield
  {title} {\bibinfo {title} {{Helicity-Resolved Raman Scattering of MoS$_2$,
  MoSe$_2$, WS$_2$, and WSe$_2$ Atomic Layers}},\ }\href
  {https://doi.org/10.1021/acs.nanolett.5b00092} {\bibfield  {journal}
  {\bibinfo  {journal} {Nano Lett.}\ }\textbf {\bibinfo {volume} {15}},\
  \bibinfo {pages} {2526–2532} (\bibinfo {year} {2015})}\BibitemShut
  {NoStop}%
\bibitem [{\citenamefont {Zhang}\ and\ \citenamefont {Niu}(2015)}]{Zhang2015}%
  \BibitemOpen
  \bibfield  {author} {\bibinfo {author} {\bibfnamefont {L.}~\bibnamefont
  {Zhang}}\ and\ \bibinfo {author} {\bibfnamefont {Q.}~\bibnamefont {Niu}},\
  }\bibfield  {title} {\bibinfo {title} {{Chiral Phonons at High-Symmetry
  Points in Monolayer Hexagonal Lattices}},\ }\href
  {https://doi.org/10.1103/PhysRevLett.115.115502} {\bibfield  {journal}
  {\bibinfo  {journal} {Phys. Rev. Lett.}\ }\textbf {\bibinfo {volume} {115}},\
  \bibinfo {pages} {115502} (\bibinfo {year} {2015})}\BibitemShut {NoStop}%
\bibitem [{\citenamefont {Zhu}\ \emph {et~al.}(2018)\citenamefont {Zhu},
  \citenamefont {Yi}, \citenamefont {Li}, \citenamefont {Xiao}, \citenamefont
  {Zhang}, \citenamefont {Yang}, \citenamefont {Kaindl}, \citenamefont {Li},
  \citenamefont {Wang},\ and\ \citenamefont {Zhang}}]{Zhu2018}%
  \BibitemOpen
  \bibfield  {author} {\bibinfo {author} {\bibfnamefont {H.}~\bibnamefont
  {Zhu}}, \bibinfo {author} {\bibfnamefont {J.}~\bibnamefont {Yi}}, \bibinfo
  {author} {\bibfnamefont {M.-Y.}\ \bibnamefont {Li}}, \bibinfo {author}
  {\bibfnamefont {J.}~\bibnamefont {Xiao}}, \bibinfo {author} {\bibfnamefont
  {L.}~\bibnamefont {Zhang}}, \bibinfo {author} {\bibfnamefont {C.-W.}\
  \bibnamefont {Yang}}, \bibinfo {author} {\bibfnamefont {R.~A.}\ \bibnamefont
  {Kaindl}}, \bibinfo {author} {\bibfnamefont {L.-J.}\ \bibnamefont {Li}},
  \bibinfo {author} {\bibfnamefont {Y.}~\bibnamefont {Wang}},\ and\ \bibinfo
  {author} {\bibfnamefont {X.}~\bibnamefont {Zhang}},\ }\bibfield  {title}
  {\bibinfo {title} {Observation of chiral phonons},\ }\href
  {https://doi.org/10.1126/science.aar2711} {\bibfield  {journal} {\bibinfo
  {journal} {Science}\ }\textbf {\bibinfo {volume} {359}},\ \bibinfo {pages}
  {579–582} (\bibinfo {year} {2018})}\BibitemShut {NoStop}%
\bibitem [{\citenamefont {Chen}\ \emph {et~al.}(2018)\citenamefont {Chen},
  \citenamefont {Zhang}, \citenamefont {Niu},\ and\ \citenamefont
  {Zhang}}]{Chen2018}%
  \BibitemOpen
  \bibfield  {author} {\bibinfo {author} {\bibfnamefont {H.}~\bibnamefont
  {Chen}}, \bibinfo {author} {\bibfnamefont {W.}~\bibnamefont {Zhang}},
  \bibinfo {author} {\bibfnamefont {Q.}~\bibnamefont {Niu}},\ and\ \bibinfo
  {author} {\bibfnamefont {L.}~\bibnamefont {Zhang}},\ }\bibfield  {title}
  {\bibinfo {title} {Chiral phonons in two-dimensional materials},\ }\href
  {https://doi.org/10.1088/2053-1583/aaf292} {\bibfield  {journal} {\bibinfo
  {journal} {2D Mater.}\ }\textbf {\bibinfo {volume} {6}},\ \bibinfo {pages}
  {012002} (\bibinfo {year} {2018})}\BibitemShut {NoStop}%
\bibitem [{\citenamefont {Chen}\ \emph {et~al.}(2019)\citenamefont {Chen},
  \citenamefont {Wu}, \citenamefont {Yang}, \citenamefont {Li},\ and\
  \citenamefont {Zhang}}]{Chen2019}%
  \BibitemOpen
  \bibfield  {author} {\bibinfo {author} {\bibfnamefont {H.}~\bibnamefont
  {Chen}}, \bibinfo {author} {\bibfnamefont {W.}~\bibnamefont {Wu}}, \bibinfo
  {author} {\bibfnamefont {S.~A.}\ \bibnamefont {Yang}}, \bibinfo {author}
  {\bibfnamefont {X.}~\bibnamefont {Li}},\ and\ \bibinfo {author}
  {\bibfnamefont {L.}~\bibnamefont {Zhang}},\ }\bibfield  {title} {\bibinfo
  {title} {Chiral phonons in kagome lattices},\ }\href
  {https://doi.org/10.1103/PhysRevB.100.094303} {\bibfield  {journal} {\bibinfo
   {journal} {Phys. Rev. B}\ }\textbf {\bibinfo {volume} {100}},\ \bibinfo
  {pages} {094303} (\bibinfo {year} {2019})}\BibitemShut {NoStop}%
\bibitem [{\citenamefont {Ptok}\ \emph {et~al.}(2021)\citenamefont {Ptok},
  \citenamefont {Kobia\l{}ka}, \citenamefont {Sternik}, \citenamefont
  {\L{}a\ifmmode~\dot{z}\else \.{z}\fi{}ewski}, \citenamefont {Jochym},
  \citenamefont {Ole\ifmmode~\acute{s}\else \'{s}\fi{}}, \citenamefont
  {Stankov},\ and\ \citenamefont {Piekarz}}]{Ptok2021}%
  \BibitemOpen
  \bibfield  {author} {\bibinfo {author} {\bibfnamefont {A.}~\bibnamefont
  {Ptok}}, \bibinfo {author} {\bibfnamefont {A.}~\bibnamefont {Kobia\l{}ka}},
  \bibinfo {author} {\bibfnamefont {M.}~\bibnamefont {Sternik}}, \bibinfo
  {author} {\bibfnamefont {J.}~\bibnamefont {\L{}a\ifmmode~\dot{z}\else
  \.{z}\fi{}ewski}}, \bibinfo {author} {\bibfnamefont {P.~T.}\ \bibnamefont
  {Jochym}}, \bibinfo {author} {\bibfnamefont {A.~M.}\ \bibnamefont
  {Ole\ifmmode~\acute{s}\else \'{s}\fi{}}}, \bibinfo {author} {\bibfnamefont
  {S.}~\bibnamefont {Stankov}},\ and\ \bibinfo {author} {\bibfnamefont
  {P.}~\bibnamefont {Piekarz}},\ }\bibfield  {title} {\bibinfo {title} {{Chiral
  phonons in the honeycomb sublattice of layered CoSn-like compounds}},\ }\href
  {https://doi.org/10.1103/PhysRevB.104.054305} {\bibfield  {journal} {\bibinfo
   {journal} {Phys. Rev. B}\ }\textbf {\bibinfo {volume} {104}},\ \bibinfo
  {pages} {054305} (\bibinfo {year} {2021})}\BibitemShut {NoStop}%
\bibitem [{\citenamefont {Saparov}\ \emph {et~al.}(2022)\citenamefont
  {Saparov}, \citenamefont {Xiong}, \citenamefont {Ren},\ and\ \citenamefont
  {Niu}}]{Saparov2022}%
  \BibitemOpen
  \bibfield  {author} {\bibinfo {author} {\bibfnamefont {D.}~\bibnamefont
  {Saparov}}, \bibinfo {author} {\bibfnamefont {B.}~\bibnamefont {Xiong}},
  \bibinfo {author} {\bibfnamefont {Y.}~\bibnamefont {Ren}},\ and\ \bibinfo
  {author} {\bibfnamefont {Q.}~\bibnamefont {Niu}},\ }\bibfield  {title}
  {\bibinfo {title} {{Lattice dynamics with molecular Berry curvature: Chiral
  optical phonons}},\ }\href {https://doi.org/10.1103/PhysRevB.105.064303}
  {\bibfield  {journal} {\bibinfo  {journal} {Phys. Rev. B}\ }\textbf {\bibinfo
  {volume} {105}},\ \bibinfo {pages} {064303} (\bibinfo {year}
  {2022})}\BibitemShut {NoStop}%
\bibitem [{\citenamefont {Chen}\ \emph {et~al.}(2024)\citenamefont {Chen},
  \citenamefont {Wang}, \citenamefont {Li},\ and\ \citenamefont
  {Zhang}}]{Chen2024}%
  \BibitemOpen
  \bibfield  {author} {\bibinfo {author} {\bibfnamefont {H.}~\bibnamefont
  {Chen}}, \bibinfo {author} {\bibfnamefont {H.}~\bibnamefont {Wang}}, \bibinfo
  {author} {\bibfnamefont {H.}~\bibnamefont {Li}},\ and\ \bibinfo {author}
  {\bibfnamefont {L.}~\bibnamefont {Zhang}},\ }\bibfield  {title} {\bibinfo
  {title} {Chiral phonons induced by adsorption on graphene},\ }\href
  {https://doi.org/10.1103/PhysRevB.110.184301} {\bibfield  {journal} {\bibinfo
   {journal} {Phys. Rev. B}\ }\textbf {\bibinfo {volume} {110}},\ \bibinfo
  {pages} {184301} (\bibinfo {year} {2024})}\BibitemShut {NoStop}%
\bibitem [{\citenamefont {Ren}\ \emph {et~al.}(2024)\citenamefont {Ren},
  \citenamefont {Bonini}, \citenamefont {Stengel}, \citenamefont {Dreyer},\
  and\ \citenamefont {Vanderbilt}}]{Ren2024PRX}%
  \BibitemOpen
  \bibfield  {author} {\bibinfo {author} {\bibfnamefont {S.}~\bibnamefont
  {Ren}}, \bibinfo {author} {\bibfnamefont {J.}~\bibnamefont {Bonini}},
  \bibinfo {author} {\bibfnamefont {M.}~\bibnamefont {Stengel}}, \bibinfo
  {author} {\bibfnamefont {C.~E.}\ \bibnamefont {Dreyer}},\ and\ \bibinfo
  {author} {\bibfnamefont {D.}~\bibnamefont {Vanderbilt}},\ }\bibfield  {title}
  {\bibinfo {title} {Adiabatic dynamics of coupled spins and phonons in
  magnetic insulators},\ }\href {https://doi.org/10.1103/PhysRevX.14.011041}
  {\bibfield  {journal} {\bibinfo  {journal} {Phys. Rev. X}\ }\textbf {\bibinfo
  {volume} {14}},\ \bibinfo {pages} {011041} (\bibinfo {year}
  {2024})}\BibitemShut {NoStop}%
\bibitem [{\citenamefont {Ishito}\ \emph {et~al.}(2022)\citenamefont {Ishito},
  \citenamefont {Mao}, \citenamefont {Kousaka}, \citenamefont {Togawa},
  \citenamefont {Iwasaki}, \citenamefont {Zhang}, \citenamefont {Murakami},
  \citenamefont {Kishine},\ and\ \citenamefont {Satoh}}]{Ishito2022}%
  \BibitemOpen
  \bibfield  {author} {\bibinfo {author} {\bibfnamefont {K.}~\bibnamefont
  {Ishito}}, \bibinfo {author} {\bibfnamefont {H.}~\bibnamefont {Mao}},
  \bibinfo {author} {\bibfnamefont {Y.}~\bibnamefont {Kousaka}}, \bibinfo
  {author} {\bibfnamefont {Y.}~\bibnamefont {Togawa}}, \bibinfo {author}
  {\bibfnamefont {S.}~\bibnamefont {Iwasaki}}, \bibinfo {author} {\bibfnamefont
  {T.}~\bibnamefont {Zhang}}, \bibinfo {author} {\bibfnamefont
  {S.}~\bibnamefont {Murakami}}, \bibinfo {author} {\bibfnamefont {J.-i.}\
  \bibnamefont {Kishine}},\ and\ \bibinfo {author} {\bibfnamefont
  {T.}~\bibnamefont {Satoh}},\ }\bibfield  {title} {\bibinfo {title} {{Truly
  chiral phonons in $\alpha$-HgS}},\ }\href
  {https://doi.org/10.1038/s41567-022-01790-x} {\bibfield  {journal} {\bibinfo
  {journal} {Nat. Phys.}\ }\textbf {\bibinfo {volume} {19}},\ \bibinfo {pages}
  {35–39} (\bibinfo {year} {2022})}\BibitemShut {NoStop}%
\bibitem [{\citenamefont {Ueda}\ \emph {et~al.}(2023)\citenamefont {Ueda},
  \citenamefont {García-Fernández}, \citenamefont {Agrestini}, \citenamefont
  {Romao}, \citenamefont {van~den Brink}, \citenamefont {Spaldin},
  \citenamefont {Zhou},\ and\ \citenamefont {Staub}}]{Ueda2023}%
  \BibitemOpen
  \bibfield  {author} {\bibinfo {author} {\bibfnamefont {H.}~\bibnamefont
  {Ueda}}, \bibinfo {author} {\bibfnamefont {M.}~\bibnamefont
  {García-Fernández}}, \bibinfo {author} {\bibfnamefont {S.}~\bibnamefont
  {Agrestini}}, \bibinfo {author} {\bibfnamefont {C.~P.}\ \bibnamefont
  {Romao}}, \bibinfo {author} {\bibfnamefont {J.}~\bibnamefont {van~den
  Brink}}, \bibinfo {author} {\bibfnamefont {N.~A.}\ \bibnamefont {Spaldin}},
  \bibinfo {author} {\bibfnamefont {K.-J.}\ \bibnamefont {Zhou}},\ and\
  \bibinfo {author} {\bibfnamefont {U.}~\bibnamefont {Staub}},\ }\bibfield
  {title} {\bibinfo {title} {Chiral phonons in quartz probed by x-rays},\
  }\href {https://doi.org/10.1038/s41586-023-06016-5} {\bibfield  {journal}
  {\bibinfo  {journal} {Nature}\ }\textbf {\bibinfo {volume} {618}},\ \bibinfo
  {pages} {946–950} (\bibinfo {year} {2023})}\BibitemShut {NoStop}%
\bibitem [{\citenamefont {Ueda}\ \emph {et~al.}(2025)\citenamefont {Ueda},
  \citenamefont {Nag}, \citenamefont {Romao}, \citenamefont
  {García-Fernández}, \citenamefont {Zhou},\ and\ \citenamefont
  {Staub}}]{Ueda2025}%
  \BibitemOpen
  \bibfield  {author} {\bibinfo {author} {\bibfnamefont {H.}~\bibnamefont
  {Ueda}}, \bibinfo {author} {\bibfnamefont {A.}~\bibnamefont {Nag}}, \bibinfo
  {author} {\bibfnamefont {C.~P.}\ \bibnamefont {Romao}}, \bibinfo {author}
  {\bibfnamefont {M.}~\bibnamefont {García-Fernández}}, \bibinfo {author}
  {\bibfnamefont {K.-J.}\ \bibnamefont {Zhou}},\ and\ \bibinfo {author}
  {\bibfnamefont {U.}~\bibnamefont {Staub}},\ }\href@noop {} {\bibinfo {title}
  {{Chiral phonons in polar LiNbO$_3$}}} (\bibinfo {year} {2025}),\ \Eprint
  {https://arxiv.org/abs/arXiv:2504.03330} {arXiv:2504.03330} \BibitemShut
  {NoStop}%
\bibitem [{\citenamefont {Yang}\ \emph {et~al.}(2026)\citenamefont {Yang},
  \citenamefont {Xiao}, \citenamefont {Mao}, \citenamefont {Li}, \citenamefont
  {Wang}, \citenamefont {Deng}, \citenamefont {Tang}, \citenamefont {Song},
  \citenamefont {Li}, \citenamefont {Yuan}, \citenamefont {Shi},\ and\
  \citenamefont {Xu}}]{Yang2026}%
  \BibitemOpen
  \bibfield  {author} {\bibinfo {author} {\bibfnamefont {Y.}~\bibnamefont
  {Yang}}, \bibinfo {author} {\bibfnamefont {Z.}~\bibnamefont {Xiao}}, \bibinfo
  {author} {\bibfnamefont {Y.}~\bibnamefont {Mao}}, \bibinfo {author}
  {\bibfnamefont {Z.}~\bibnamefont {Li}}, \bibinfo {author} {\bibfnamefont
  {Z.}~\bibnamefont {Wang}}, \bibinfo {author} {\bibfnamefont {T.}~\bibnamefont
  {Deng}}, \bibinfo {author} {\bibfnamefont {Y.}~\bibnamefont {Tang}}, \bibinfo
  {author} {\bibfnamefont {Z.-D.}\ \bibnamefont {Song}}, \bibinfo {author}
  {\bibfnamefont {Y.}~\bibnamefont {Li}}, \bibinfo {author} {\bibfnamefont
  {H.}~\bibnamefont {Yuan}}, \bibinfo {author} {\bibfnamefont {M.}~\bibnamefont
  {Shi}},\ and\ \bibinfo {author} {\bibfnamefont {Y.}~\bibnamefont {Xu}},\
  }\bibfield  {title} {\bibinfo {title} {Symmetry-guided catalogue of chiral
  phonon materials},\ }\href {https://doi.org/10.1038/s41567-026-03260-0}
  {\bibfield  {journal} {\bibinfo  {journal} {Nat. Phys.}\ }\textbf {\bibinfo
  {volume} {22}},\ \bibinfo {pages} {03260} (\bibinfo {year}
  {2026})}\BibitemShut {NoStop}%
\bibitem [{\citenamefont {Juraschek}\ \emph {et~al.}(2025)\citenamefont
  {Juraschek}, \citenamefont {Geilhufe}, \citenamefont {Zhu}, \citenamefont
  {Basini}, \citenamefont {Baum}, \citenamefont {Baydin}, \citenamefont
  {Chaudhary}, \citenamefont {Fechner}, \citenamefont {Flebus}, \citenamefont
  {Grissonnanche}, \citenamefont {Kirilyuk}, \citenamefont {Lemeshko},
  \citenamefont {Maehrlein}, \citenamefont {Mignolet}, \citenamefont
  {Murakami}, \citenamefont {Niu}, \citenamefont {Nowak}, \citenamefont
  {Romao}, \citenamefont {Rostami}, \citenamefont {Satoh}, \citenamefont
  {Spaldin}, \citenamefont {Ueda},\ and\ \citenamefont
  {Zhang}}]{Juraschek2025}%
  \BibitemOpen
  \bibfield  {author} {\bibinfo {author} {\bibfnamefont {D.~M.}\ \bibnamefont
  {Juraschek}}, \bibinfo {author} {\bibfnamefont {R.~M.}\ \bibnamefont
  {Geilhufe}}, \bibinfo {author} {\bibfnamefont {H.}~\bibnamefont {Zhu}},
  \bibinfo {author} {\bibfnamefont {M.}~\bibnamefont {Basini}}, \bibinfo
  {author} {\bibfnamefont {P.}~\bibnamefont {Baum}}, \bibinfo {author}
  {\bibfnamefont {A.}~\bibnamefont {Baydin}}, \bibinfo {author} {\bibfnamefont
  {S.}~\bibnamefont {Chaudhary}}, \bibinfo {author} {\bibfnamefont
  {M.}~\bibnamefont {Fechner}}, \bibinfo {author} {\bibfnamefont
  {B.}~\bibnamefont {Flebus}}, \bibinfo {author} {\bibfnamefont
  {G.}~\bibnamefont {Grissonnanche}}, \bibinfo {author} {\bibfnamefont {A.~I.}\
  \bibnamefont {Kirilyuk}}, \bibinfo {author} {\bibfnamefont {M.}~\bibnamefont
  {Lemeshko}}, \bibinfo {author} {\bibfnamefont {S.~F.}\ \bibnamefont
  {Maehrlein}}, \bibinfo {author} {\bibfnamefont {M.}~\bibnamefont {Mignolet}},
  \bibinfo {author} {\bibfnamefont {S.}~\bibnamefont {Murakami}}, \bibinfo
  {author} {\bibfnamefont {Q.}~\bibnamefont {Niu}}, \bibinfo {author}
  {\bibfnamefont {U.}~\bibnamefont {Nowak}}, \bibinfo {author} {\bibfnamefont
  {C.~P.}\ \bibnamefont {Romao}}, \bibinfo {author} {\bibfnamefont
  {H.}~\bibnamefont {Rostami}}, \bibinfo {author} {\bibfnamefont
  {T.}~\bibnamefont {Satoh}}, \bibinfo {author} {\bibfnamefont {N.~A.}\
  \bibnamefont {Spaldin}}, \bibinfo {author} {\bibfnamefont {H.}~\bibnamefont
  {Ueda}},\ and\ \bibinfo {author} {\bibfnamefont {L.}~\bibnamefont {Zhang}},\
  }\bibfield  {title} {\bibinfo {title} {Chiral phonons},\ }\href
  {https://doi.org/10.1038/s41567-025-03001-9} {\bibfield  {journal} {\bibinfo
  {journal} {Nat. Phys.}\ }\textbf {\bibinfo {volume} {21}},\ \bibinfo {pages}
  {1532–1540} (\bibinfo {year} {2025})}\BibitemShut {NoStop}%
\bibitem [{\citenamefont {Wang}\ \emph
  {et~al.}(2025{\natexlab{a}})\citenamefont {Wang}, \citenamefont {Xu},
  \citenamefont {Liu}, \citenamefont {Wang}, \citenamefont {Zhang},\ and\
  \citenamefont {Zhang}}]{Wang2025alteraxial}%
  \BibitemOpen
  \bibfield  {author} {\bibinfo {author} {\bibfnamefont {F.}~\bibnamefont
  {Wang}}, \bibinfo {author} {\bibfnamefont {J.}~\bibnamefont {Xu}}, \bibinfo
  {author} {\bibfnamefont {X.}~\bibnamefont {Liu}}, \bibinfo {author}
  {\bibfnamefont {H.}~\bibnamefont {Wang}}, \bibinfo {author} {\bibfnamefont
  {L.}~\bibnamefont {Zhang}},\ and\ \bibinfo {author} {\bibfnamefont
  {H.}~\bibnamefont {Zhang}},\ }\href@noop {} {\bibinfo {title} {Alteraxial
  phonons in collinear magnets}} (\bibinfo {year} {2025}{\natexlab{a}}),\
  \Eprint {https://arxiv.org/abs/arXiv:2512.07518} {arXiv:2512.07518}
  \BibitemShut {NoStop}%
\bibitem [{\citenamefont {Zhang}\ \emph
  {et~al.}(2023{\natexlab{a}})\citenamefont {Zhang}, \citenamefont {Huang},
  \citenamefont {Pan}, \citenamefont {Du}, \citenamefont {Zhang},\ and\
  \citenamefont {Murakami}}]{Zhang2023Weylphonon}%
  \BibitemOpen
  \bibfield  {author} {\bibinfo {author} {\bibfnamefont {T.}~\bibnamefont
  {Zhang}}, \bibinfo {author} {\bibfnamefont {Z.}~\bibnamefont {Huang}},
  \bibinfo {author} {\bibfnamefont {Z.}~\bibnamefont {Pan}}, \bibinfo {author}
  {\bibfnamefont {L.}~\bibnamefont {Du}}, \bibinfo {author} {\bibfnamefont
  {G.}~\bibnamefont {Zhang}},\ and\ \bibinfo {author} {\bibfnamefont
  {S.}~\bibnamefont {Murakami}},\ }\bibfield  {title} {\bibinfo {title} {Weyl
  phonons in chiral crystals},\ }\href
  {https://doi.org/10.1021/acs.nanolett.3c02132} {\bibfield  {journal}
  {\bibinfo  {journal} {Nano Lett.}\ }\textbf {\bibinfo {volume} {23}},\
  \bibinfo {pages} {7561} (\bibinfo {year} {2023}{\natexlab{a}})}\BibitemShut
  {NoStop}%
\bibitem [{\citenamefont {Ma}\ \emph {et~al.}(2024)\citenamefont {Ma},
  \citenamefont {Wang},\ and\ \citenamefont {Chen}}]{Ma2024}%
  \BibitemOpen
  \bibfield  {author} {\bibinfo {author} {\bibfnamefont {B.}~\bibnamefont
  {Ma}}, \bibinfo {author} {\bibfnamefont {Z.~D.}\ \bibnamefont {Wang}},\ and\
  \bibinfo {author} {\bibfnamefont {G.~v.}\ \bibnamefont {Chen}},\ }\bibfield
  {title} {\bibinfo {title} {{Chiral Phonons Induced from Spin Dynamics via
  Magnetoelastic Anisotropy}},\ }\href
  {https://doi.org/10.1103/PhysRevLett.133.246604} {\bibfield  {journal}
  {\bibinfo  {journal} {Phys. Rev. Lett.}\ }\textbf {\bibinfo {volume} {133}},\
  \bibinfo {pages} {246604} (\bibinfo {year} {2024})}\BibitemShut {NoStop}%
\bibitem [{\citenamefont {Wei\ss{}enhofer}\ \emph
  {et~al.}(2025{\natexlab{a}})\citenamefont {Wei\ss{}enhofer}, \citenamefont
  {Rieger}, \citenamefont {Mrudul}, \citenamefont {Mikadze}, \citenamefont
  {Nowak},\ and\ \citenamefont {Oppeneer}}]{Weissenhofer2025}%
  \BibitemOpen
  \bibfield  {author} {\bibinfo {author} {\bibfnamefont {M.}~\bibnamefont
  {Wei\ss{}enhofer}}, \bibinfo {author} {\bibfnamefont {P.}~\bibnamefont
  {Rieger}}, \bibinfo {author} {\bibfnamefont {M.~S.}\ \bibnamefont {Mrudul}},
  \bibinfo {author} {\bibfnamefont {L.}~\bibnamefont {Mikadze}}, \bibinfo
  {author} {\bibfnamefont {U.}~\bibnamefont {Nowak}},\ and\ \bibinfo {author}
  {\bibfnamefont {P.~M.}\ \bibnamefont {Oppeneer}},\ }\bibfield  {title}
  {\bibinfo {title} {{Chiral Phonons Arising from Chirality-Selective
  Magnon-Phonon Coupling}},\ }\href {https://doi.org/10.1103/j7bs-2zbx}
  {\bibfield  {journal} {\bibinfo  {journal} {Phys. Rev. Lett.}\ }\textbf
  {\bibinfo {volume} {135}},\ \bibinfo {pages} {216701} (\bibinfo {year}
  {2025}{\natexlab{a}})}\BibitemShut {NoStop}%
\bibitem [{\citenamefont {Che}\ \emph {et~al.}(2025)\citenamefont {Che},
  \citenamefont {Liang}, \citenamefont {Cui}, \citenamefont {Li}, \citenamefont
  {Lu}, \citenamefont {Sang}, \citenamefont {Li}, \citenamefont {Dong},
  \citenamefont {Zhao}, \citenamefont {Zhang}, \citenamefont {Sun},
  \citenamefont {Jiang}, \citenamefont {Liu}, \citenamefont {Jin},
  \citenamefont {Zhang},\ and\ \citenamefont {Yang}}]{Che2025}%
  \BibitemOpen
  \bibfield  {author} {\bibinfo {author} {\bibfnamefont {M.}~\bibnamefont
  {Che}}, \bibinfo {author} {\bibfnamefont {J.}~\bibnamefont {Liang}}, \bibinfo
  {author} {\bibfnamefont {Y.}~\bibnamefont {Cui}}, \bibinfo {author}
  {\bibfnamefont {H.}~\bibnamefont {Li}}, \bibinfo {author} {\bibfnamefont
  {B.}~\bibnamefont {Lu}}, \bibinfo {author} {\bibfnamefont {W.}~\bibnamefont
  {Sang}}, \bibinfo {author} {\bibfnamefont {X.}~\bibnamefont {Li}}, \bibinfo
  {author} {\bibfnamefont {X.}~\bibnamefont {Dong}}, \bibinfo {author}
  {\bibfnamefont {L.}~\bibnamefont {Zhao}}, \bibinfo {author} {\bibfnamefont
  {S.}~\bibnamefont {Zhang}}, \bibinfo {author} {\bibfnamefont
  {T.}~\bibnamefont {Sun}}, \bibinfo {author} {\bibfnamefont {W.}~\bibnamefont
  {Jiang}}, \bibinfo {author} {\bibfnamefont {E.}~\bibnamefont {Liu}}, \bibinfo
  {author} {\bibfnamefont {F.}~\bibnamefont {Jin}}, \bibinfo {author}
  {\bibfnamefont {T.}~\bibnamefont {Zhang}},\ and\ \bibinfo {author}
  {\bibfnamefont {L.}~\bibnamefont {Yang}},\ }\bibfield  {title} {\bibinfo
  {title} {{Magnetic Order Induced Chiral Phonons in a Ferromagnetic Weyl
  Semimetal}},\ }\href {https://doi.org/10.1103/PhysRevLett.134.196906}
  {\bibfield  {journal} {\bibinfo  {journal} {Phys. Rev. Lett.}\ }\textbf
  {\bibinfo {volume} {134}},\ \bibinfo {pages} {196906} (\bibinfo {year}
  {2025})}\BibitemShut {NoStop}%
\bibitem [{\citenamefont {Streib}(2021)}]{Streib2021}%
  \BibitemOpen
  \bibfield  {author} {\bibinfo {author} {\bibfnamefont {S.}~\bibnamefont
  {Streib}},\ }\bibfield  {title} {\bibinfo {title} {Difference between angular
  momentum and pseudoangular momentum},\ }\href
  {https://doi.org/10.1103/PhysRevB.103.L100409} {\bibfield  {journal}
  {\bibinfo  {journal} {Phys. Rev. B}\ }\textbf {\bibinfo {volume} {103}},\
  \bibinfo {pages} {L100409} (\bibinfo {year} {2021})}\BibitemShut {NoStop}%
\bibitem [{\citenamefont {Zhang}\ and\ \citenamefont
  {Murakami}(2022)}]{Zhang2022chiralpam}%
  \BibitemOpen
  \bibfield  {author} {\bibinfo {author} {\bibfnamefont {T.}~\bibnamefont
  {Zhang}}\ and\ \bibinfo {author} {\bibfnamefont {S.}~\bibnamefont
  {Murakami}},\ }\bibfield  {title} {\bibinfo {title} {Chiral phonons and
  pseudoangular momentum in nonsymmorphic systems},\ }\href
  {https://doi.org/10.1103/PhysRevResearch.4.L012024} {\bibfield  {journal}
  {\bibinfo  {journal} {Phys. Rev. Res.}\ }\textbf {\bibinfo {volume} {4}},\
  \bibinfo {pages} {L012024} (\bibinfo {year} {2022})}\BibitemShut {NoStop}%
\bibitem [{\citenamefont {Wang}\ \emph {et~al.}(2024)\citenamefont {Wang},
  \citenamefont {Sun}, \citenamefont {Li},\ and\ \citenamefont
  {Zhang}}]{Wang2024}%
  \BibitemOpen
  \bibfield  {author} {\bibinfo {author} {\bibfnamefont {T.}~\bibnamefont
  {Wang}}, \bibinfo {author} {\bibfnamefont {H.}~\bibnamefont {Sun}}, \bibinfo
  {author} {\bibfnamefont {X.}~\bibnamefont {Li}},\ and\ \bibinfo {author}
  {\bibfnamefont {L.}~\bibnamefont {Zhang}},\ }\bibfield  {title} {\bibinfo
  {title} {Chiral phonons: Prediction, verification, and application},\ }\href
  {https://doi.org/10.1021/acs.nanolett.4c00606} {\bibfield  {journal}
  {\bibinfo  {journal} {Nano Lett.}\ }\textbf {\bibinfo {volume} {24}},\
  \bibinfo {pages} {4311} (\bibinfo {year} {2024})}\BibitemShut {NoStop}%
\bibitem [{\citenamefont {Chen}\ \emph {et~al.}(2025)\citenamefont {Chen},
  \citenamefont {Qin}, \citenamefont {Zhang}, \citenamefont {Cui},
  \citenamefont {Niu},\ and\ \citenamefont {Zhang}}]{Chen2025}%
  \BibitemOpen
  \bibfield  {author} {\bibinfo {author} {\bibfnamefont {Y.}~\bibnamefont
  {Chen}}, \bibinfo {author} {\bibfnamefont {W.}~\bibnamefont {Qin}}, \bibinfo
  {author} {\bibfnamefont {S.}~\bibnamefont {Zhang}}, \bibinfo {author}
  {\bibfnamefont {P.}~\bibnamefont {Cui}}, \bibinfo {author} {\bibfnamefont
  {Q.}~\bibnamefont {Niu}},\ and\ \bibinfo {author} {\bibfnamefont
  {Z.}~\bibnamefont {Zhang}},\ }\bibfield  {title} {\bibinfo {title}
  {{Emergence of Chiral Phonons in Two-Dimensional Kagome Lattices Harboring
  Electronic Chirality}},\ }\href {https://doi.org/10.1103/bfll-sdrb}
  {\bibfield  {journal} {\bibinfo  {journal} {Phys. Rev. Lett.}\ }\textbf
  {\bibinfo {volume} {135}},\ \bibinfo {pages} {126608} (\bibinfo {year}
  {2025})}\BibitemShut {NoStop}%
\bibitem [{\citenamefont {Liu}\ \emph {et~al.}(2017)\citenamefont {Liu},
  \citenamefont {Lian}, \citenamefont {Li}, \citenamefont {Xu},\ and\
  \citenamefont {Duan}}]{Liu2017}%
  \BibitemOpen
  \bibfield  {author} {\bibinfo {author} {\bibfnamefont {Y.}~\bibnamefont
  {Liu}}, \bibinfo {author} {\bibfnamefont {C.-S.}\ \bibnamefont {Lian}},
  \bibinfo {author} {\bibfnamefont {Y.}~\bibnamefont {Li}}, \bibinfo {author}
  {\bibfnamefont {Y.}~\bibnamefont {Xu}},\ and\ \bibinfo {author}
  {\bibfnamefont {W.}~\bibnamefont {Duan}},\ }\bibfield  {title} {\bibinfo
  {title} {{Pseudospins and Topological Effects of Phonons in a Kekul\'e
  Lattice}},\ }\href {https://doi.org/10.1103/PhysRevLett.119.255901}
  {\bibfield  {journal} {\bibinfo  {journal} {Phys. Rev. Lett.}\ }\textbf
  {\bibinfo {volume} {119}},\ \bibinfo {pages} {255901} (\bibinfo {year}
  {2017})}\BibitemShut {NoStop}%
\bibitem [{\citenamefont {Li}\ \emph {et~al.}(2020)\citenamefont {Li},
  \citenamefont {Fauqu\'e}, \citenamefont {Zhu},\ and\ \citenamefont
  {Behnia}}]{Li2020}%
  \BibitemOpen
  \bibfield  {author} {\bibinfo {author} {\bibfnamefont {X.}~\bibnamefont
  {Li}}, \bibinfo {author} {\bibfnamefont {B.}~\bibnamefont {Fauqu\'e}},
  \bibinfo {author} {\bibfnamefont {Z.}~\bibnamefont {Zhu}},\ and\ \bibinfo
  {author} {\bibfnamefont {K.}~\bibnamefont {Behnia}},\ }\bibfield  {title}
  {\bibinfo {title} {{Phonon Thermal Hall Effect in Strontium Titanate}},\
  }\href {https://doi.org/10.1103/PhysRevLett.124.105901} {\bibfield  {journal}
  {\bibinfo  {journal} {Phys. Rev. Lett.}\ }\textbf {\bibinfo {volume} {124}},\
  \bibinfo {pages} {105901} (\bibinfo {year} {2020})}\BibitemShut {NoStop}%
\bibitem [{\citenamefont {Park}\ and\ \citenamefont {Yang}(2020)}]{Park2020}%
  \BibitemOpen
  \bibfield  {author} {\bibinfo {author} {\bibfnamefont {S.}~\bibnamefont
  {Park}}\ and\ \bibinfo {author} {\bibfnamefont {B.-J.}\ \bibnamefont
  {Yang}},\ }\bibfield  {title} {\bibinfo {title} {{Phonon Angular Momentum
  Hall Effect}},\ }\href {https://doi.org/10.1021/acs.nanolett.0c03220}
  {\bibfield  {journal} {\bibinfo  {journal} {Nano Lett.}\ }\textbf {\bibinfo
  {volume} {20}},\ \bibinfo {pages} {7694–7699} (\bibinfo {year}
  {2020})}\BibitemShut {NoStop}%
\bibitem [{\citenamefont {Sun}\ \emph {et~al.}(2021)\citenamefont {Sun},
  \citenamefont {Gao},\ and\ \citenamefont {Wang}}]{Sun2021}%
  \BibitemOpen
  \bibfield  {author} {\bibinfo {author} {\bibfnamefont {K.}~\bibnamefont
  {Sun}}, \bibinfo {author} {\bibfnamefont {Z.}~\bibnamefont {Gao}},\ and\
  \bibinfo {author} {\bibfnamefont {J.-S.}\ \bibnamefont {Wang}},\ }\bibfield
  {title} {\bibinfo {title} {{Phonon Hall effect with first-principles
  calculations}},\ }\href {https://doi.org/10.1103/PhysRevB.103.214301}
  {\bibfield  {journal} {\bibinfo  {journal} {Phys. Rev. B}\ }\textbf {\bibinfo
  {volume} {103}},\ \bibinfo {pages} {214301} (\bibinfo {year}
  {2021})}\BibitemShut {NoStop}%
\bibitem [{\citenamefont {Oh}\ and\ \citenamefont {Nagaosa}(2025)}]{Oh2025}%
  \BibitemOpen
  \bibfield  {author} {\bibinfo {author} {\bibfnamefont {T.}~\bibnamefont
  {Oh}}\ and\ \bibinfo {author} {\bibfnamefont {N.}~\bibnamefont {Nagaosa}},\
  }\bibfield  {title} {\bibinfo {title} {{Phonon Thermal Hall Effect in Mott
  Insulators via Skew Scattering by the Scalar Spin Chirality}},\ }\href
  {https://doi.org/10.1103/PhysRevX.15.011036} {\bibfield  {journal} {\bibinfo
  {journal} {Phys. Rev. X}\ }\textbf {\bibinfo {volume} {15}},\ \bibinfo
  {pages} {011036} (\bibinfo {year} {2025})}\BibitemShut {NoStop}%
\bibitem [{\citenamefont {Kim}\ \emph {et~al.}(2023)\citenamefont {Kim},
  \citenamefont {Vetter}, \citenamefont {Yan}, \citenamefont {Yang},
  \citenamefont {Wang}, \citenamefont {Sun}, \citenamefont {Yang},
  \citenamefont {Comstock}, \citenamefont {Li}, \citenamefont {Zhou},
  \citenamefont {Zhang}, \citenamefont {You}, \citenamefont {Sun},\ and\
  \citenamefont {Liu}}]{Kim2023}%
  \BibitemOpen
  \bibfield  {author} {\bibinfo {author} {\bibfnamefont {K.}~\bibnamefont
  {Kim}}, \bibinfo {author} {\bibfnamefont {E.}~\bibnamefont {Vetter}},
  \bibinfo {author} {\bibfnamefont {L.}~\bibnamefont {Yan}}, \bibinfo {author}
  {\bibfnamefont {C.}~\bibnamefont {Yang}}, \bibinfo {author} {\bibfnamefont
  {Z.}~\bibnamefont {Wang}}, \bibinfo {author} {\bibfnamefont {R.}~\bibnamefont
  {Sun}}, \bibinfo {author} {\bibfnamefont {Y.}~\bibnamefont {Yang}}, \bibinfo
  {author} {\bibfnamefont {A.~H.}\ \bibnamefont {Comstock}}, \bibinfo {author}
  {\bibfnamefont {X.}~\bibnamefont {Li}}, \bibinfo {author} {\bibfnamefont
  {J.}~\bibnamefont {Zhou}}, \bibinfo {author} {\bibfnamefont {L.}~\bibnamefont
  {Zhang}}, \bibinfo {author} {\bibfnamefont {W.}~\bibnamefont {You}}, \bibinfo
  {author} {\bibfnamefont {D.}~\bibnamefont {Sun}},\ and\ \bibinfo {author}
  {\bibfnamefont {J.}~\bibnamefont {Liu}},\ }\bibfield  {title} {\bibinfo
  {title} {{Chiral-phonon-activated spin Seebeck effect}},\ }\href
  {https://doi.org/10.1038/s41563-023-01473-9} {\bibfield  {journal} {\bibinfo
  {journal} {Nat. Mater.}\ }\textbf {\bibinfo {volume} {22}},\ \bibinfo {pages}
  {322–328} (\bibinfo {year} {2023})}\BibitemShut {NoStop}%
\bibitem [{\citenamefont {Yang}\ \emph {et~al.}(2024)\citenamefont {Yang},
  \citenamefont {Yu}, \citenamefont {Sui}, \citenamefont {Ge}, \citenamefont
  {Jin}, \citenamefont {Liu}, \citenamefont {Chen}, \citenamefont {Qi},\ and\
  \citenamefont {Yue}}]{Yang2024}%
  \BibitemOpen
  \bibfield  {author} {\bibinfo {author} {\bibfnamefont {S.}~\bibnamefont
  {Yang}}, \bibinfo {author} {\bibfnamefont {Y.}~\bibnamefont {Yu}}, \bibinfo
  {author} {\bibfnamefont {F.}~\bibnamefont {Sui}}, \bibinfo {author}
  {\bibfnamefont {R.}~\bibnamefont {Ge}}, \bibinfo {author} {\bibfnamefont
  {R.}~\bibnamefont {Jin}}, \bibinfo {author} {\bibfnamefont {B.}~\bibnamefont
  {Liu}}, \bibinfo {author} {\bibfnamefont {Y.}~\bibnamefont {Chen}}, \bibinfo
  {author} {\bibfnamefont {R.}~\bibnamefont {Qi}},\ and\ \bibinfo {author}
  {\bibfnamefont {F.}~\bibnamefont {Yue}},\ }\bibfield  {title} {\bibinfo
  {title} {{Chiral Phonon, Valley Polarization, and Inter/Intravalley
  Scattering in a van der Waals ReSe$_2$ Semiconductor}},\ }\href
  {https://doi.org/10.1021/acsnano.4c15485} {\bibfield  {journal} {\bibinfo
  {journal} {ACS Nano}\ }\textbf {\bibinfo {volume} {18}},\ \bibinfo {pages}
  {33754–33764} (\bibinfo {year} {2024})}\BibitemShut {NoStop}%
\bibitem [{\citenamefont {Wei\ss{}enhofer}\ \emph
  {et~al.}(2025{\natexlab{b}})\citenamefont {Wei\ss{}enhofer}, \citenamefont
  {Rieger}, \citenamefont {Mrudul}, \citenamefont {Mikadze}, \citenamefont
  {Nowak},\ and\ \citenamefont {Oppeneer}}]{Whofer2025}%
  \BibitemOpen
  \bibfield  {author} {\bibinfo {author} {\bibfnamefont {M.}~\bibnamefont
  {Wei\ss{}enhofer}}, \bibinfo {author} {\bibfnamefont {P.}~\bibnamefont
  {Rieger}}, \bibinfo {author} {\bibfnamefont {M.~S.}\ \bibnamefont {Mrudul}},
  \bibinfo {author} {\bibfnamefont {L.}~\bibnamefont {Mikadze}}, \bibinfo
  {author} {\bibfnamefont {U.}~\bibnamefont {Nowak}},\ and\ \bibinfo {author}
  {\bibfnamefont {P.~M.}\ \bibnamefont {Oppeneer}},\ }\bibfield  {title}
  {\bibinfo {title} {{Chiral Phonons Arising from Chirality-Selective
  Magnon-Phonon Coupling}},\ }\href {https://doi.org/10.1103/j7bs-2zbx}
  {\bibfield  {journal} {\bibinfo  {journal} {Phys. Rev. Lett.}\ }\textbf
  {\bibinfo {volume} {135}},\ \bibinfo {pages} {216701} (\bibinfo {year}
  {2025}{\natexlab{b}})}\BibitemShut {NoStop}%
\bibitem [{\citenamefont {Yaniv}\ and\ \citenamefont
  {Juraschek}(2025)}]{yaniv2025multicolor}%
  \BibitemOpen
  \bibfield  {author} {\bibinfo {author} {\bibfnamefont {O.}~\bibnamefont
  {Yaniv}}\ and\ \bibinfo {author} {\bibfnamefont {D.~M.}\ \bibnamefont
  {Juraschek}},\ }\bibfield  {title} {\bibinfo {title} {Multicolor phonon
  excitation in terahertz cavities},\ }\href
  {https://doi.org/10.1103/fwr7-6jj6} {\bibfield  {journal} {\bibinfo
  {journal} {Phys. Rev. Lett.}\ }\textbf {\bibinfo {volume} {135}},\ \bibinfo
  {pages} {246901} (\bibinfo {year} {2025})}\BibitemShut {NoStop}%
\bibitem [{\citenamefont {Juraschek}\ and\ \citenamefont
  {Spaldin}(2019)}]{Juraschek2019}%
  \BibitemOpen
  \bibfield  {author} {\bibinfo {author} {\bibfnamefont {D.~M.}\ \bibnamefont
  {Juraschek}}\ and\ \bibinfo {author} {\bibfnamefont {N.~A.}\ \bibnamefont
  {Spaldin}},\ }\bibfield  {title} {\bibinfo {title} {Orbital magnetic moments
  of phonons},\ }\href {https://doi.org/10.1103/PhysRevMaterials.3.064405}
  {\bibfield  {journal} {\bibinfo  {journal} {Phys. Rev. Mater.}\ }\textbf
  {\bibinfo {volume} {3}},\ \bibinfo {pages} {064405} (\bibinfo {year}
  {2019})}\BibitemShut {NoStop}%
\bibitem [{\citenamefont {Luo}\ \emph {et~al.}(2023)\citenamefont {Luo},
  \citenamefont {Lin}, \citenamefont {Zhang}, \citenamefont {Chen},
  \citenamefont {Blackert}, \citenamefont {Xu}, \citenamefont {Yakobson},\ and\
  \citenamefont {Zhu}}]{Luo2023}%
  \BibitemOpen
  \bibfield  {author} {\bibinfo {author} {\bibfnamefont {J.}~\bibnamefont
  {Luo}}, \bibinfo {author} {\bibfnamefont {T.}~\bibnamefont {Lin}}, \bibinfo
  {author} {\bibfnamefont {J.}~\bibnamefont {Zhang}}, \bibinfo {author}
  {\bibfnamefont {X.}~\bibnamefont {Chen}}, \bibinfo {author} {\bibfnamefont
  {E.~R.}\ \bibnamefont {Blackert}}, \bibinfo {author} {\bibfnamefont
  {R.}~\bibnamefont {Xu}}, \bibinfo {author} {\bibfnamefont {B.~I.}\
  \bibnamefont {Yakobson}},\ and\ \bibinfo {author} {\bibfnamefont
  {H.}~\bibnamefont {Zhu}},\ }\bibfield  {title} {\bibinfo {title} {Large
  effective magnetic fields from chiral phonons in rare-earth halides},\ }\href
  {https://doi.org/10.1126/science.adi9601} {\bibfield  {journal} {\bibinfo
  {journal} {Science}\ }\textbf {\bibinfo {volume} {382}},\ \bibinfo {pages}
  {698–702} (\bibinfo {year} {2023})}\BibitemShut {NoStop}%
\bibitem [{\citenamefont {Wu}\ \emph {et~al.}(2025)\citenamefont {Wu},
  \citenamefont {Zhou}, \citenamefont {Bao}, \citenamefont {Li}, \citenamefont
  {Wen}, \citenamefont {Wan},\ and\ \citenamefont {Zhang}}]{Wu2025}%
  \BibitemOpen
  \bibfield  {author} {\bibinfo {author} {\bibfnamefont {F.}~\bibnamefont
  {Wu}}, \bibinfo {author} {\bibfnamefont {J.}~\bibnamefont {Zhou}}, \bibinfo
  {author} {\bibfnamefont {S.}~\bibnamefont {Bao}}, \bibinfo {author}
  {\bibfnamefont {L.}~\bibnamefont {Li}}, \bibinfo {author} {\bibfnamefont
  {J.}~\bibnamefont {Wen}}, \bibinfo {author} {\bibfnamefont {Y.}~\bibnamefont
  {Wan}},\ and\ \bibinfo {author} {\bibfnamefont {Q.}~\bibnamefont {Zhang}},\
  }\bibfield  {title} {\bibinfo {title} {{Magnetic Switching of Phonon Angular
  Momentum in a Ferrimagnetic Insulator}},\ }\href
  {https://doi.org/10.1103/gcgl-9sbb} {\bibfield  {journal} {\bibinfo
  {journal} {Phys. Rev. Lett.}\ }\textbf {\bibinfo {volume} {134}},\ \bibinfo
  {pages} {236701} (\bibinfo {year} {2025})}\BibitemShut {NoStop}%
\bibitem [{\citenamefont {Yang}\ \emph {et~al.}(2025)\citenamefont {Yang},
  \citenamefont {Zhu}, \citenamefont {Steigleder}, \citenamefont {Liu},
  \citenamefont {Liu}, \citenamefont {Qiu}, \citenamefont {Zhang},\ and\
  \citenamefont {Dressel}}]{Yang2025}%
  \BibitemOpen
  \bibfield  {author} {\bibinfo {author} {\bibfnamefont {R.}~\bibnamefont
  {Yang}}, \bibinfo {author} {\bibfnamefont {Y.-Y.}\ \bibnamefont {Zhu}},
  \bibinfo {author} {\bibfnamefont {M.}~\bibnamefont {Steigleder}}, \bibinfo
  {author} {\bibfnamefont {Y.-C.}\ \bibnamefont {Liu}}, \bibinfo {author}
  {\bibfnamefont {C.-C.}\ \bibnamefont {Liu}}, \bibinfo {author} {\bibfnamefont
  {X.-G.}\ \bibnamefont {Qiu}}, \bibinfo {author} {\bibfnamefont
  {T.}~\bibnamefont {Zhang}},\ and\ \bibinfo {author} {\bibfnamefont
  {M.}~\bibnamefont {Dressel}},\ }\bibfield  {title} {\bibinfo {title}
  {{Inherent Circular Dichroism of Phonons in Magnetic Weyl Semimetal
  ${\mathrm{Co}}_{3}{\text{Sn}}_{2}{\mathrm{S}}_{2}$}},\ }\href
  {https://doi.org/10.1103/PhysRevLett.134.196905} {\bibfield  {journal}
  {\bibinfo  {journal} {Phys. Rev. Lett.}\ }\textbf {\bibinfo {volume} {134}},\
  \bibinfo {pages} {196905} (\bibinfo {year} {2025})}\BibitemShut {NoStop}%
\bibitem [{\citenamefont {Nova}\ \emph {et~al.}(2016)\citenamefont {Nova},
  \citenamefont {Cartella}, \citenamefont {Cantaluppi}, \citenamefont
  {F\"{o}rst}, \citenamefont {Bossini}, \citenamefont {Mikhaylovskiy},
  \citenamefont {Kimel}, \citenamefont {Merlin},\ and\ \citenamefont
  {Cavalleri}}]{Nova2016}%
  \BibitemOpen
  \bibfield  {author} {\bibinfo {author} {\bibfnamefont {T.~F.}\ \bibnamefont
  {Nova}}, \bibinfo {author} {\bibfnamefont {A.}~\bibnamefont {Cartella}},
  \bibinfo {author} {\bibfnamefont {A.}~\bibnamefont {Cantaluppi}}, \bibinfo
  {author} {\bibfnamefont {M.}~\bibnamefont {F\"{o}rst}}, \bibinfo {author}
  {\bibfnamefont {D.}~\bibnamefont {Bossini}}, \bibinfo {author} {\bibfnamefont
  {R.~V.}\ \bibnamefont {Mikhaylovskiy}}, \bibinfo {author} {\bibfnamefont
  {A.~V.}\ \bibnamefont {Kimel}}, \bibinfo {author} {\bibfnamefont
  {R.}~\bibnamefont {Merlin}},\ and\ \bibinfo {author} {\bibfnamefont
  {A.}~\bibnamefont {Cavalleri}},\ }\bibfield  {title} {\bibinfo {title} {An
  effective magnetic field from optically driven phonons},\ }\href
  {https://doi.org/10.1038/nphys3925} {\bibfield  {journal} {\bibinfo
  {journal} {Nat. Phys.}\ }\textbf {\bibinfo {volume} {13}},\ \bibinfo {pages}
  {132–136} (\bibinfo {year} {2016})}\BibitemShut {NoStop}%
\bibitem [{\citenamefont {Ren}\ \emph {et~al.}(2021)\citenamefont {Ren},
  \citenamefont {Xiao}, \citenamefont {Saparov},\ and\ \citenamefont
  {Niu}}]{Ren2021}%
  \BibitemOpen
  \bibfield  {author} {\bibinfo {author} {\bibfnamefont {Y.}~\bibnamefont
  {Ren}}, \bibinfo {author} {\bibfnamefont {C.}~\bibnamefont {Xiao}}, \bibinfo
  {author} {\bibfnamefont {D.}~\bibnamefont {Saparov}},\ and\ \bibinfo {author}
  {\bibfnamefont {Q.}~\bibnamefont {Niu}},\ }\bibfield  {title} {\bibinfo
  {title} {{Phonon Magnetic Moment from Electronic Topological
  Magnetization}},\ }\href {https://doi.org/10.1103/PhysRevLett.127.186403}
  {\bibfield  {journal} {\bibinfo  {journal} {Phys. Rev. Lett.}\ }\textbf
  {\bibinfo {volume} {127}},\ \bibinfo {pages} {186403} (\bibinfo {year}
  {2021})}\BibitemShut {NoStop}%
\bibitem [{\citenamefont {Juraschek}\ \emph {et~al.}(2022)\citenamefont
  {Juraschek}, \citenamefont {Neuman},\ and\ \citenamefont
  {Narang}}]{Juraschek2022}%
  \BibitemOpen
  \bibfield  {author} {\bibinfo {author} {\bibfnamefont {D.~M.}\ \bibnamefont
  {Juraschek}}, \bibinfo {author} {\bibfnamefont {T.~c.~v.}\ \bibnamefont
  {Neuman}},\ and\ \bibinfo {author} {\bibfnamefont {P.}~\bibnamefont
  {Narang}},\ }\bibfield  {title} {\bibinfo {title} {Giant effective magnetic
  fields from optically driven chiral phonons in $4f$ paramagnets},\ }\href
  {https://doi.org/10.1103/PhysRevResearch.4.013129} {\bibfield  {journal}
  {\bibinfo  {journal} {Phys. Rev. Res.}\ }\textbf {\bibinfo {volume} {4}},\
  \bibinfo {pages} {013129} (\bibinfo {year} {2022})}\BibitemShut {NoStop}%
\bibitem [{\citenamefont {Geilhufe}\ and\ \citenamefont
  {Hergert}(2023)}]{Geilhufe2023}%
  \BibitemOpen
  \bibfield  {author} {\bibinfo {author} {\bibfnamefont {R.~M.}\ \bibnamefont
  {Geilhufe}}\ and\ \bibinfo {author} {\bibfnamefont {W.}~\bibnamefont
  {Hergert}},\ }\bibfield  {title} {\bibinfo {title} {{Electron magnetic moment
  of transient chiral phonons in ${\mathrm{KTaO}}_{3}$}},\ }\href
  {https://doi.org/10.1103/PhysRevB.107.L020406} {\bibfield  {journal}
  {\bibinfo  {journal} {Phys. Rev. B}\ }\textbf {\bibinfo {volume} {107}},\
  \bibinfo {pages} {L020406} (\bibinfo {year} {2023})}\BibitemShut {NoStop}%
\bibitem [{\citenamefont {Zhang}\ \emph
  {et~al.}(2023{\natexlab{b}})\citenamefont {Zhang}, \citenamefont {Ren},
  \citenamefont {Wang}, \citenamefont {Cao},\ and\ \citenamefont
  {Xiao}}]{Zhang2023}%
  \BibitemOpen
  \bibfield  {author} {\bibinfo {author} {\bibfnamefont {X.-W.}\ \bibnamefont
  {Zhang}}, \bibinfo {author} {\bibfnamefont {Y.}~\bibnamefont {Ren}}, \bibinfo
  {author} {\bibfnamefont {C.}~\bibnamefont {Wang}}, \bibinfo {author}
  {\bibfnamefont {T.}~\bibnamefont {Cao}},\ and\ \bibinfo {author}
  {\bibfnamefont {D.}~\bibnamefont {Xiao}},\ }\bibfield  {title} {\bibinfo
  {title} {{Gate-Tunable Phonon Magnetic Moment in Bilayer Graphene}},\ }\href
  {https://doi.org/10.1103/PhysRevLett.130.226302} {\bibfield  {journal}
  {\bibinfo  {journal} {Phys. Rev. Lett.}\ }\textbf {\bibinfo {volume} {130}},\
  \bibinfo {pages} {226302} (\bibinfo {year} {2023}{\natexlab{b}})}\BibitemShut
  {NoStop}%
\bibitem [{\citenamefont {Chaudhary}\ \emph {et~al.}(2024)\citenamefont
  {Chaudhary}, \citenamefont {Juraschek}, \citenamefont {Rodriguez-Vega},\ and\
  \citenamefont {Fiete}}]{Chaudhury2024}%
  \BibitemOpen
  \bibfield  {author} {\bibinfo {author} {\bibfnamefont {S.}~\bibnamefont
  {Chaudhary}}, \bibinfo {author} {\bibfnamefont {D.~M.}\ \bibnamefont
  {Juraschek}}, \bibinfo {author} {\bibfnamefont {M.}~\bibnamefont
  {Rodriguez-Vega}},\ and\ \bibinfo {author} {\bibfnamefont {G.~A.}\
  \bibnamefont {Fiete}},\ }\bibfield  {title} {\bibinfo {title} {Giant
  effective magnetic moments of chiral phonons from orbit-lattice coupling},\
  }\href {https://doi.org/10.1103/PhysRevB.110.094401} {\bibfield  {journal}
  {\bibinfo  {journal} {Phys. Rev. B}\ }\textbf {\bibinfo {volume} {110}},\
  \bibinfo {pages} {094401} (\bibinfo {year} {2024})}\BibitemShut {NoStop}%
\bibitem [{\citenamefont {Yao}\ \emph {et~al.}(2025)\citenamefont {Yao},
  \citenamefont {Go}, \citenamefont {Mokrousov},\ and\ \citenamefont
  {Murakami}}]{Yao2025}%
  \BibitemOpen
  \bibfield  {author} {\bibinfo {author} {\bibfnamefont {D.}~\bibnamefont
  {Yao}}, \bibinfo {author} {\bibfnamefont {D.}~\bibnamefont {Go}}, \bibinfo
  {author} {\bibfnamefont {Y.}~\bibnamefont {Mokrousov}},\ and\ \bibinfo
  {author} {\bibfnamefont {S.}~\bibnamefont {Murakami}},\ }\href@noop {}
  {\bibinfo {title} {Dynamical orbital angular momentum induced by circularly
  polarized phonons}} (\bibinfo {year} {2025}),\ \Eprint
  {https://arxiv.org/abs/arXiv:2511.09271} {arXiv:2511.09271} \BibitemShut
  {NoStop}%
\bibitem [{\citenamefont {Coh}(2023)}]{Coh2023}%
  \BibitemOpen
  \bibfield  {author} {\bibinfo {author} {\bibfnamefont {S.}~\bibnamefont
  {Coh}},\ }\bibfield  {title} {\bibinfo {title} {Classification of materials
  with phonon angular momentum and microscopic origin of angular momentum},\
  }\href {https://doi.org/10.1103/PhysRevB.108.134307} {\bibfield  {journal}
  {\bibinfo  {journal} {Phys. Rev. B}\ }\textbf {\bibinfo {volume} {108}},\
  \bibinfo {pages} {134307} (\bibinfo {year} {2023})}\BibitemShut {NoStop}%
\bibitem [{\citenamefont {Chaudhary}\ \emph {et~al.}(2026)\citenamefont
  {Chaudhary}, \citenamefont {Romao},\ and\ \citenamefont
  {Juraschek}}]{chaudhary2026anomalous}%
  \BibitemOpen
  \bibfield  {author} {\bibinfo {author} {\bibfnamefont {S.}~\bibnamefont
  {Chaudhary}}, \bibinfo {author} {\bibfnamefont {C.~P.}\ \bibnamefont
  {Romao}},\ and\ \bibinfo {author} {\bibfnamefont {D.~M.}\ \bibnamefont
  {Juraschek}},\ }\bibfield  {title} {\bibinfo {title} {Anomalous phonon
  magnetic moments},\ }\href {https://doi.org/10.1103/xzgb-33w5} {\bibfield
  {journal} {\bibinfo  {journal} {Phys. Rev. Res.}\ }\textbf {\bibinfo {volume}
  {8}},\ \bibinfo {pages} {023053} (\bibinfo {year} {2026})}\BibitemShut
  {NoStop}%
\bibitem [{\citenamefont {Bonini}\ \emph {et~al.}(2023)\citenamefont {Bonini},
  \citenamefont {Ren}, \citenamefont {Vanderbilt}, \citenamefont {Stengel},
  \citenamefont {Dreyer},\ and\ \citenamefont {Coh}}]{Bonini2023}%
  \BibitemOpen
  \bibfield  {author} {\bibinfo {author} {\bibfnamefont {J.}~\bibnamefont
  {Bonini}}, \bibinfo {author} {\bibfnamefont {S.}~\bibnamefont {Ren}},
  \bibinfo {author} {\bibfnamefont {D.}~\bibnamefont {Vanderbilt}}, \bibinfo
  {author} {\bibfnamefont {M.}~\bibnamefont {Stengel}}, \bibinfo {author}
  {\bibfnamefont {C.~E.}\ \bibnamefont {Dreyer}},\ and\ \bibinfo {author}
  {\bibfnamefont {S.}~\bibnamefont {Coh}},\ }\bibfield  {title} {\bibinfo
  {title} {{Frequency Splitting of Chiral Phonons from Broken Time-Reversal
  Symmetry in ${\mathrm{CrI}}_{3}$}},\ }\href
  {https://doi.org/10.1103/PhysRevLett.130.086701} {\bibfield  {journal}
  {\bibinfo  {journal} {Phys. Rev. Lett.}\ }\textbf {\bibinfo {volume} {130}},\
  \bibinfo {pages} {086701} (\bibinfo {year} {2023})}\BibitemShut {NoStop}%
\bibitem [{\citenamefont {Wang}\ \emph
  {et~al.}(2025{\natexlab{b}})\citenamefont {Wang}, \citenamefont {Liu},
  \citenamefont {Sun}, \citenamefont {Wang}, \citenamefont {Murakami},
  \citenamefont {Zhang}, \citenamefont {Zhang},\ and\ \citenamefont
  {Xing}}]{Wang2025ab}%
  \BibitemOpen
  \bibfield  {author} {\bibinfo {author} {\bibfnamefont {F.}~\bibnamefont
  {Wang}}, \bibinfo {author} {\bibfnamefont {X.}~\bibnamefont {Liu}}, \bibinfo
  {author} {\bibfnamefont {H.}~\bibnamefont {Sun}}, \bibinfo {author}
  {\bibfnamefont {H.}~\bibnamefont {Wang}}, \bibinfo {author} {\bibfnamefont
  {S.}~\bibnamefont {Murakami}}, \bibinfo {author} {\bibfnamefont
  {L.}~\bibnamefont {Zhang}}, \bibinfo {author} {\bibfnamefont
  {H.}~\bibnamefont {Zhang}},\ and\ \bibinfo {author} {\bibfnamefont
  {D.}~\bibnamefont {Xing}},\ }\bibfield  {title} {\bibinfo {title} {Ab initio
  theory of phonon magnetic moment induced by electron-phonon coupling in
  magnetic materials},\ }\href {https://doi.org/10.1103/tpjd-dh1m} {\bibfield
  {journal} {\bibinfo  {journal} {Phys. Rev. Lett.}\ }\textbf {\bibinfo
  {volume} {135}},\ \bibinfo {pages} {256701} (\bibinfo {year}
  {2025}{\natexlab{b}})}\BibitemShut {NoStop}%
\bibitem [{\citenamefont {Fennie}\ and\ \citenamefont
  {Rabe}(2006)}]{fennie2006magnetically}%
  \BibitemOpen
  \bibfield  {author} {\bibinfo {author} {\bibfnamefont {C.~J.}\ \bibnamefont
  {Fennie}}\ and\ \bibinfo {author} {\bibfnamefont {K.~M.}\ \bibnamefont
  {Rabe}},\ }\bibfield  {title} {\bibinfo {title} {{Magnetically Induced Phonon
  Anisotropy in ${\mathrm{ZnCr}}_{2}{\mathrm{O}}_{4}$ from First Principles}},\
  }\href {https://doi.org/10.1103/PhysRevLett.96.205505} {\bibfield  {journal}
  {\bibinfo  {journal} {Phys. Rev. Lett.}\ }\textbf {\bibinfo {volume} {96}},\
  \bibinfo {pages} {205505} (\bibinfo {year} {2006})}\BibitemShut {NoStop}%
\bibitem [{\citenamefont {Kant}\ \emph {et~al.}(2012)\citenamefont {Kant},
  \citenamefont {Schmidt}, \citenamefont {Wang}, \citenamefont {Mayr},
  \citenamefont {Tsurkan}, \citenamefont {Deisenhofer},\ and\ \citenamefont
  {Loidl}}]{kant2012universal}%
  \BibitemOpen
  \bibfield  {author} {\bibinfo {author} {\bibfnamefont {C.}~\bibnamefont
  {Kant}}, \bibinfo {author} {\bibfnamefont {M.}~\bibnamefont {Schmidt}},
  \bibinfo {author} {\bibfnamefont {Z.}~\bibnamefont {Wang}}, \bibinfo {author}
  {\bibfnamefont {F.}~\bibnamefont {Mayr}}, \bibinfo {author} {\bibfnamefont
  {V.}~\bibnamefont {Tsurkan}}, \bibinfo {author} {\bibfnamefont
  {J.}~\bibnamefont {Deisenhofer}},\ and\ \bibinfo {author} {\bibfnamefont
  {A.}~\bibnamefont {Loidl}},\ }\bibfield  {title} {\bibinfo {title} {Universal
  exchange-driven phonon splitting in antiferromagnets},\ }\href
  {https://doi.org/10.1103/PhysRevLett.108.177203} {\bibfield  {journal}
  {\bibinfo  {journal} {Phys. Rev. Lett.}\ }\textbf {\bibinfo {volume} {108}},\
  \bibinfo {pages} {177203} (\bibinfo {year} {2012})}\BibitemShut {NoStop}%
\bibitem [{\citenamefont {Sohn}\ \emph {et~al.}(2017)\citenamefont {Sohn},
  \citenamefont {Kim}, \citenamefont {Sandilands}, \citenamefont {Hien},
  \citenamefont {Kim}, \citenamefont {Park}, \citenamefont {Kim}, \citenamefont
  {Moon}, \citenamefont {Yamaura}, \citenamefont {Hiroi},\ and\ \citenamefont
  {Noh}}]{sohn2017strong}%
  \BibitemOpen
  \bibfield  {author} {\bibinfo {author} {\bibfnamefont {C.~H.}\ \bibnamefont
  {Sohn}}, \bibinfo {author} {\bibfnamefont {C.~H.}\ \bibnamefont {Kim}},
  \bibinfo {author} {\bibfnamefont {L.~J.}\ \bibnamefont {Sandilands}},
  \bibinfo {author} {\bibfnamefont {N.~T.~M.}\ \bibnamefont {Hien}}, \bibinfo
  {author} {\bibfnamefont {S.~Y.}\ \bibnamefont {Kim}}, \bibinfo {author}
  {\bibfnamefont {H.~J.}\ \bibnamefont {Park}}, \bibinfo {author}
  {\bibfnamefont {K.~W.}\ \bibnamefont {Kim}}, \bibinfo {author} {\bibfnamefont
  {S.~J.}\ \bibnamefont {Moon}}, \bibinfo {author} {\bibfnamefont
  {J.}~\bibnamefont {Yamaura}}, \bibinfo {author} {\bibfnamefont
  {Z.}~\bibnamefont {Hiroi}},\ and\ \bibinfo {author} {\bibfnamefont {T.~W.}\
  \bibnamefont {Noh}},\ }\bibfield  {title} {\bibinfo {title} {{Strong
  Spin-Phonon Coupling Mediated by Single Ion Anisotropy in the All-In--All-Out
  Pyrochlore Magnet ${\mathrm{Cd}}_{2}{\mathrm{Os}}_{2}{\mathrm{O}}_{7}$}},\
  }\href {https://doi.org/10.1103/PhysRevLett.118.117201} {\bibfield  {journal}
  {\bibinfo  {journal} {Phys. Rev. Lett.}\ }\textbf {\bibinfo {volume} {118}},\
  \bibinfo {pages} {117201} (\bibinfo {year} {2017})}\BibitemShut {NoStop}%
\bibitem [{\citenamefont {Son}\ \emph {et~al.}(2019)\citenamefont {Son},
  \citenamefont {Park}, \citenamefont {Kim}, \citenamefont {Cho}, \citenamefont
  {Kim}, \citenamefont {Sandilands}, \citenamefont {Sohn}, \citenamefont
  {Park}, \citenamefont {Moon},\ and\ \citenamefont
  {Noh}}]{son2019unconventional}%
  \BibitemOpen
  \bibfield  {author} {\bibinfo {author} {\bibfnamefont {J.}~\bibnamefont
  {Son}}, \bibinfo {author} {\bibfnamefont {B.~C.}\ \bibnamefont {Park}},
  \bibinfo {author} {\bibfnamefont {C.~H.}\ \bibnamefont {Kim}}, \bibinfo
  {author} {\bibfnamefont {H.}~\bibnamefont {Cho}}, \bibinfo {author}
  {\bibfnamefont {S.~Y.}\ \bibnamefont {Kim}}, \bibinfo {author} {\bibfnamefont
  {L.~J.}\ \bibnamefont {Sandilands}}, \bibinfo {author} {\bibfnamefont
  {C.}~\bibnamefont {Sohn}}, \bibinfo {author} {\bibfnamefont {J.-G.}\
  \bibnamefont {Park}}, \bibinfo {author} {\bibfnamefont {S.~J.}\ \bibnamefont
  {Moon}},\ and\ \bibinfo {author} {\bibfnamefont {T.~W.}\ \bibnamefont
  {Noh}},\ }\bibfield  {title} {\bibinfo {title} {Unconventional spin-phonon
  coupling via the dzyaloshinskii–moriya interaction},\ }\bibfield  {journal}
  {\bibinfo  {journal} {npj Quantum Mater.}\ }\textbf {\bibinfo {volume} {4}},\
  \href {https://doi.org/10.1038/s41535-019-0157-0} {10.1038/s41535-019-0157-0}
  (\bibinfo {year} {2019})\BibitemShut {NoStop}%
\bibitem [{\citenamefont {Cui}\ \emph {et~al.}(2023)\citenamefont {Cui},
  \citenamefont {Bostr\"{o}m}, \citenamefont {Ozerov}, \citenamefont {Wu},
  \citenamefont {Jiang}, \citenamefont {Chu}, \citenamefont {Li}, \citenamefont
  {Liu}, \citenamefont {Xu}, \citenamefont {Rubio},\ and\ \citenamefont
  {Zhang}}]{Cui2023}%
  \BibitemOpen
  \bibfield  {author} {\bibinfo {author} {\bibfnamefont {J.}~\bibnamefont
  {Cui}}, \bibinfo {author} {\bibfnamefont {E.~V.}\ \bibnamefont
  {Bostr\"{o}m}}, \bibinfo {author} {\bibfnamefont {M.}~\bibnamefont {Ozerov}},
  \bibinfo {author} {\bibfnamefont {F.}~\bibnamefont {Wu}}, \bibinfo {author}
  {\bibfnamefont {Q.}~\bibnamefont {Jiang}}, \bibinfo {author} {\bibfnamefont
  {J.-H.}\ \bibnamefont {Chu}}, \bibinfo {author} {\bibfnamefont
  {C.}~\bibnamefont {Li}}, \bibinfo {author} {\bibfnamefont {F.}~\bibnamefont
  {Liu}}, \bibinfo {author} {\bibfnamefont {X.}~\bibnamefont {Xu}}, \bibinfo
  {author} {\bibfnamefont {A.}~\bibnamefont {Rubio}},\ and\ \bibinfo {author}
  {\bibfnamefont {Q.}~\bibnamefont {Zhang}},\ }\bibfield  {title} {\bibinfo
  {title} {Chirality selective magnon--phonon hybridization and magnon-induced
  chiral phonons in a layered zigzag antiferromagnet},\ }\href
  {https://doi.org/10.1038/s41467-023-39123-y} {\bibfield  {journal} {\bibinfo
  {journal} {Nat. Commun.}\ }\textbf {\bibinfo {volume} {14}},\ \bibinfo
  {pages} {39123} (\bibinfo {year} {2023})}\BibitemShut {NoStop}%
\bibitem [{\citenamefont {Mead}\ and\ \citenamefont
  {Truhlar}(1979)}]{Mead1979}%
  \BibitemOpen
  \bibfield  {author} {\bibinfo {author} {\bibfnamefont {C.~A.}\ \bibnamefont
  {Mead}}\ and\ \bibinfo {author} {\bibfnamefont {D.~G.}\ \bibnamefont
  {Truhlar}},\ }\bibfield  {title} {\bibinfo {title} {{On the determination of
  Born–Oppenheimer nuclear motion wave functions including complications due
  to conical intersections and identical nuclei}},\ }\href
  {https://doi.org/10.1063/1.437734} {\bibfield  {journal} {\bibinfo  {journal}
  {J. Chem. Phys.}\ }\textbf {\bibinfo {volume} {70}},\ \bibinfo {pages}
  {2284–2296} (\bibinfo {year} {1979})}\BibitemShut {NoStop}%
\bibitem [{\citenamefont {Xiao}\ \emph {et~al.}(2010)\citenamefont {Xiao},
  \citenamefont {Chang},\ and\ \citenamefont {Niu}}]{Xiao2010}%
  \BibitemOpen
  \bibfield  {author} {\bibinfo {author} {\bibfnamefont {D.}~\bibnamefont
  {Xiao}}, \bibinfo {author} {\bibfnamefont {M.-C.}\ \bibnamefont {Chang}},\
  and\ \bibinfo {author} {\bibfnamefont {Q.}~\bibnamefont {Niu}},\ }\bibfield
  {title} {\bibinfo {title} {Berry phase effects on electronic properties},\
  }\href {https://doi.org/10.1103/RevModPhys.82.1959} {\bibfield  {journal}
  {\bibinfo  {journal} {Rev. Mod. Phys.}\ }\textbf {\bibinfo {volume} {82}},\
  \bibinfo {pages} {1959} (\bibinfo {year} {2010})}\BibitemShut {NoStop}%
\bibitem [{\citenamefont {Zhang}\ \emph {et~al.}(2026)\citenamefont {Zhang},
  \citenamefont {Wang}, \citenamefont {Zhang},\ and\ \citenamefont
  {Zhang}}]{zhang2026general}%
  \BibitemOpen
  \bibfield  {author} {\bibinfo {author} {\bibfnamefont {S.}~\bibnamefont
  {Zhang}}, \bibinfo {author} {\bibfnamefont {M.}~\bibnamefont {Wang}},
  \bibinfo {author} {\bibfnamefont {P.}~\bibnamefont {Zhang}},\ and\ \bibinfo
  {author} {\bibfnamefont {T.}~\bibnamefont {Zhang}},\ }\bibfield  {title}
  {\bibinfo {title} {General ab initio framework for electronic-order–induced
  lattice-dynamics symmetry breaking},\ }\bibfield  {journal} {\bibinfo
  {journal} {Sci. Adv.}\ }\textbf {\bibinfo {volume} {12}},\ \href
  {https://doi.org/10.1126/sciadv.aed7081} {10.1126/sciadv.aed7081} (\bibinfo
  {year} {2026})\BibitemShut {NoStop}%
\bibitem [{\citenamefont {Ren}\ \emph {et~al.}(2025)\citenamefont {Ren},
  \citenamefont {Saparov},\ and\ \citenamefont {Niu}}]{Ren2025}%
  \BibitemOpen
  \bibfield  {author} {\bibinfo {author} {\bibfnamefont {Y.}~\bibnamefont
  {Ren}}, \bibinfo {author} {\bibfnamefont {D.}~\bibnamefont {Saparov}},\ and\
  \bibinfo {author} {\bibfnamefont {Q.}~\bibnamefont {Niu}},\ }\bibfield
  {title} {\bibinfo {title} {{Nonreciprocal Phonons in
  $\mathcal{P}\mathcal{T}$-Symmetric Antiferromagnets}},\ }\href
  {https://doi.org/10.1103/PhysRevLett.134.206701} {\bibfield  {journal}
  {\bibinfo  {journal} {Phys. Rev. Lett.}\ }\textbf {\bibinfo {volume} {134}},\
  \bibinfo {pages} {206701} (\bibinfo {year} {2025})}\BibitemShut {NoStop}%
\bibitem [{Das()}]{DasSM}%
  \BibitemOpen
  \href@noop {} {}\bibinfo {note} {See Supplemental Material for detailed
  calculations of the lattice dynamics, molecular Berry curvature, symmetry
  transformation of the AFCP modes, and rigorous proofs of the symmetry
  constraints on the gyroscopic coupling.}\BibitemShut {Stop}%
\bibitem [{\citenamefont {Liu}\ \emph {et~al.}(2011)\citenamefont {Liu},
  \citenamefont {Jiang},\ and\ \citenamefont {Yao}}]{liu2011low}%
  \BibitemOpen
  \bibfield  {author} {\bibinfo {author} {\bibfnamefont {C.-C.}\ \bibnamefont
  {Liu}}, \bibinfo {author} {\bibfnamefont {H.}~\bibnamefont {Jiang}},\ and\
  \bibinfo {author} {\bibfnamefont {Y.}~\bibnamefont {Yao}},\ }\bibfield
  {title} {\bibinfo {title} {Low-energy effective hamiltonian involving
  spin-orbit coupling in silicene and two-dimensional germanium and tin},\
  }\href {https://doi.org/10.1103/PhysRevB.84.195430} {\bibfield  {journal}
  {\bibinfo  {journal} {Phys. Rev. B}\ }\textbf {\bibinfo {volume} {84}},\
  \bibinfo {pages} {195430} (\bibinfo {year} {2011})}\BibitemShut {NoStop}%
\bibitem [{\citenamefont {McCann}\ and\ \citenamefont
  {Koshino}(2013)}]{mccann2013electronic}%
  \BibitemOpen
  \bibfield  {author} {\bibinfo {author} {\bibfnamefont {E.}~\bibnamefont
  {McCann}}\ and\ \bibinfo {author} {\bibfnamefont {M.}~\bibnamefont
  {Koshino}},\ }\bibfield  {title} {\bibinfo {title} {The electronic properties
  of bilayer graphene},\ }\href {https://doi.org/10.1088/0034-4885/76/5/056503}
  {\bibfield  {journal} {\bibinfo  {journal} {Rep. Prog. Phys.}\ }\textbf
  {\bibinfo {volume} {76}},\ \bibinfo {pages} {056503} (\bibinfo {year}
  {2013})}\BibitemShut {NoStop}%
\bibitem [{\citenamefont {Wu}\ \emph {et~al.}(2022)\citenamefont {Wu},
  \citenamefont {Yao}, \citenamefont {Lin}, \citenamefont {R\"{o}sner},
  \citenamefont {Du}, \citenamefont {Watanabe}, \citenamefont {Taniguchi},
  \citenamefont {Tan}, \citenamefont {Haas},\ and\ \citenamefont
  {Wang}}]{Wu2022}%
  \BibitemOpen
  \bibfield  {author} {\bibinfo {author} {\bibfnamefont {J.}~\bibnamefont
  {Wu}}, \bibinfo {author} {\bibfnamefont {Y.}~\bibnamefont {Yao}}, \bibinfo
  {author} {\bibfnamefont {M.-L.}\ \bibnamefont {Lin}}, \bibinfo {author}
  {\bibfnamefont {M.}~\bibnamefont {R\"{o}sner}}, \bibinfo {author}
  {\bibfnamefont {Z.}~\bibnamefont {Du}}, \bibinfo {author} {\bibfnamefont
  {K.}~\bibnamefont {Watanabe}}, \bibinfo {author} {\bibfnamefont
  {T.}~\bibnamefont {Taniguchi}}, \bibinfo {author} {\bibfnamefont {P.-H.}\
  \bibnamefont {Tan}}, \bibinfo {author} {\bibfnamefont {S.}~\bibnamefont
  {Haas}},\ and\ \bibinfo {author} {\bibfnamefont {H.}~\bibnamefont {Wang}},\
  }\bibfield  {title} {\bibinfo {title} {Spin--phonon coupling in ferromagnetic
  monolayer chromium tribromide},\ }\href
  {https://doi.org/10.1002/adma.202108506} {\bibfield  {journal} {\bibinfo
  {journal} {Adv. Mater.}\ }\textbf {\bibinfo {volume} {34}},\ \bibinfo {pages}
  {2108506} (\bibinfo {year} {2022})}\BibitemShut {NoStop}%
\bibitem [{\citenamefont {Zhang}\ \emph {et~al.}(2025)\citenamefont {Zhang},
  \citenamefont {Wu}, \citenamefont {Yang}, \citenamefont {Jin}, \citenamefont
  {Xiao}, \citenamefont {Zhang},\ and\ \citenamefont
  {Chang}}]{Zhang2025spinphonon}%
  \BibitemOpen
  \bibfield  {author} {\bibinfo {author} {\bibfnamefont {G.}~\bibnamefont
  {Zhang}}, \bibinfo {author} {\bibfnamefont {H.}~\bibnamefont {Wu}}, \bibinfo
  {author} {\bibfnamefont {L.}~\bibnamefont {Yang}}, \bibinfo {author}
  {\bibfnamefont {W.}~\bibnamefont {Jin}}, \bibinfo {author} {\bibfnamefont
  {B.}~\bibnamefont {Xiao}}, \bibinfo {author} {\bibfnamefont {W.}~\bibnamefont
  {Zhang}},\ and\ \bibinfo {author} {\bibfnamefont {H.}~\bibnamefont {Chang}},\
  }\bibfield  {title} {\bibinfo {title} {{Lattice Vibration, Raman Modes and
  Room-Temperature Spin-Phonon Coupling in Intrinsic Two-Dimensional van der
  Waals Ferromagnetic Fe$_3$GaTe$_2$}},\ }\href
  {https://doi.org/10.1021/acsmaterialslett.4c02526} {\bibfield  {journal}
  {\bibinfo  {journal} {ACS Mater. Lett.}\ }\textbf {\bibinfo {volume} {7}},\
  \bibinfo {pages} {1289} (\bibinfo {year} {2025})}\BibitemShut {NoStop}%
\bibitem [{\citenamefont {Kargar}\ \emph {et~al.}(2020)\citenamefont {Kargar},
  \citenamefont {Coleman}, \citenamefont {Ghosh}, \citenamefont {Lee},
  \citenamefont {Gomez}, \citenamefont {Liu}, \citenamefont {Magana},
  \citenamefont {Barani}, \citenamefont {Mohammadzadeh}, \citenamefont
  {Debnath}, \citenamefont {Wilson}, \citenamefont {Lake},\ and\ \citenamefont
  {Balandin}}]{Kargar2020}%
  \BibitemOpen
  \bibfield  {author} {\bibinfo {author} {\bibfnamefont {F.}~\bibnamefont
  {Kargar}}, \bibinfo {author} {\bibfnamefont {E.~A.}\ \bibnamefont {Coleman}},
  \bibinfo {author} {\bibfnamefont {S.}~\bibnamefont {Ghosh}}, \bibinfo
  {author} {\bibfnamefont {J.}~\bibnamefont {Lee}}, \bibinfo {author}
  {\bibfnamefont {M.~J.}\ \bibnamefont {Gomez}}, \bibinfo {author}
  {\bibfnamefont {Y.}~\bibnamefont {Liu}}, \bibinfo {author} {\bibfnamefont
  {A.~S.}\ \bibnamefont {Magana}}, \bibinfo {author} {\bibfnamefont
  {Z.}~\bibnamefont {Barani}}, \bibinfo {author} {\bibfnamefont
  {A.}~\bibnamefont {Mohammadzadeh}}, \bibinfo {author} {\bibfnamefont
  {B.}~\bibnamefont {Debnath}}, \bibinfo {author} {\bibfnamefont {R.~B.}\
  \bibnamefont {Wilson}}, \bibinfo {author} {\bibfnamefont {R.~K.}\
  \bibnamefont {Lake}},\ and\ \bibinfo {author} {\bibfnamefont {A.~A.}\
  \bibnamefont {Balandin}},\ }\bibfield  {title} {\bibinfo {title} {{Phonon and
  Thermal Properties of Quasi-Two-Dimensional FePS$_3$ and MnPS$_3$
  Antiferromagnetic Semiconductors}},\ }\href
  {https://doi.org/10.1021/acsnano.9b09839} {\bibfield  {journal} {\bibinfo
  {journal} {ACS Nano}\ }\textbf {\bibinfo {volume} {14}},\ \bibinfo {pages}
  {2424} (\bibinfo {year} {2020})}\BibitemShut {NoStop}%
\bibitem [{\citenamefont {Dixit}\ \emph {et~al.}(2026)\citenamefont {Dixit},
  \citenamefont {Yang}, \citenamefont {Li}, \citenamefont {Chen}, \citenamefont
  {Jia},\ and\ \citenamefont {Wang}}]{Dixit2026}%
  \BibitemOpen
  \bibfield  {author} {\bibinfo {author} {\bibfnamefont {B.}~\bibnamefont
  {Dixit}}, \bibinfo {author} {\bibfnamefont {Y.}~\bibnamefont {Yang}},
  \bibinfo {author} {\bibfnamefont {S.}~\bibnamefont {Li}}, \bibinfo {author}
  {\bibfnamefont {Y.-C.}\ \bibnamefont {Chen}}, \bibinfo {author}
  {\bibfnamefont {Q.}~\bibnamefont {Jia}},\ and\ \bibinfo {author}
  {\bibfnamefont {J.-P.}\ \bibnamefont {Wang}},\ }\bibfield  {title} {\bibinfo
  {title} {2d magnetic and topological quantum materials and devices for
  ultralow power spintronics},\ }\href {https://doi.org/10.1002/adfm.202528026}
  {\bibfield  {journal} {\bibinfo  {journal} {Adv. Funct. Mater.}\ }\textbf
  {\bibinfo {volume} {36}},\ \bibinfo {pages} {2528026} (\bibinfo {year}
  {2026})}\BibitemShut {NoStop}%
\bibitem [{\citenamefont {Otrokov}\ \emph {et~al.}(2019)\citenamefont
  {Otrokov}, \citenamefont {Klimovskikh}, \citenamefont {Bentmann},
  \citenamefont {Estyunin}, \citenamefont {Zeugner}, \citenamefont {Aliev},
  \citenamefont {Gaß}, \citenamefont {Wolter}, \citenamefont {Koroleva},
  \citenamefont {Shikin}, \citenamefont {Blanco-Rey}, \citenamefont {Hoffmann},
  \citenamefont {Rusinov}, \citenamefont {Vyazovskaya}, \citenamefont
  {Eremeev}, \citenamefont {Koroteev}, \citenamefont {Kuznetsov}, \citenamefont
  {Freyse}, \citenamefont {Sánchez-Barriga}, \citenamefont {Amiraslanov},
  \citenamefont {Babanly}, \citenamefont {Mamedov}, \citenamefont {Abdullayev},
  \citenamefont {Zverev}, \citenamefont {Alfonsov}, \citenamefont {Kataev},
  \citenamefont {B\"{u}chner}, \citenamefont {Schwier}, \citenamefont {Kumar},
  \citenamefont {Kimura}, \citenamefont {Petaccia}, \citenamefont {Di~Santo},
  \citenamefont {Vidal}, \citenamefont {Schatz}, \citenamefont {Kißner},
  \citenamefont {\"{U}nzelmann}, \citenamefont {Min}, \citenamefont {Moser},
  \citenamefont {Peixoto}, \citenamefont {Reinert}, \citenamefont {Ernst},
  \citenamefont {Echenique}, \citenamefont {Isaeva},\ and\ \citenamefont
  {Chulkov}}]{otrokov2019prediction}%
  \BibitemOpen
  \bibfield  {author} {\bibinfo {author} {\bibfnamefont {M.~M.}\ \bibnamefont
  {Otrokov}}, \bibinfo {author} {\bibfnamefont {I.~I.}\ \bibnamefont
  {Klimovskikh}}, \bibinfo {author} {\bibfnamefont {H.}~\bibnamefont
  {Bentmann}}, \bibinfo {author} {\bibfnamefont {D.}~\bibnamefont {Estyunin}},
  \bibinfo {author} {\bibfnamefont {A.}~\bibnamefont {Zeugner}}, \bibinfo
  {author} {\bibfnamefont {Z.~S.}\ \bibnamefont {Aliev}}, \bibinfo {author}
  {\bibfnamefont {S.}~\bibnamefont {Gaß}}, \bibinfo {author} {\bibfnamefont
  {A.~U.~B.}\ \bibnamefont {Wolter}}, \bibinfo {author} {\bibfnamefont {A.~V.}\
  \bibnamefont {Koroleva}}, \bibinfo {author} {\bibfnamefont {A.~M.}\
  \bibnamefont {Shikin}}, \bibinfo {author} {\bibfnamefont {M.}~\bibnamefont
  {Blanco-Rey}}, \bibinfo {author} {\bibfnamefont {M.}~\bibnamefont
  {Hoffmann}}, \bibinfo {author} {\bibfnamefont {I.~P.}\ \bibnamefont
  {Rusinov}}, \bibinfo {author} {\bibfnamefont {A.~Y.}\ \bibnamefont
  {Vyazovskaya}}, \bibinfo {author} {\bibfnamefont {S.~V.}\ \bibnamefont
  {Eremeev}}, \bibinfo {author} {\bibfnamefont {Y.~M.}\ \bibnamefont
  {Koroteev}}, \bibinfo {author} {\bibfnamefont {V.~M.}\ \bibnamefont
  {Kuznetsov}}, \bibinfo {author} {\bibfnamefont {F.}~\bibnamefont {Freyse}},
  \bibinfo {author} {\bibfnamefont {J.}~\bibnamefont {Sánchez-Barriga}},
  \bibinfo {author} {\bibfnamefont {I.~R.}\ \bibnamefont {Amiraslanov}},
  \bibinfo {author} {\bibfnamefont {M.~B.}\ \bibnamefont {Babanly}}, \bibinfo
  {author} {\bibfnamefont {N.~T.}\ \bibnamefont {Mamedov}}, \bibinfo {author}
  {\bibfnamefont {N.~A.}\ \bibnamefont {Abdullayev}}, \bibinfo {author}
  {\bibfnamefont {V.~N.}\ \bibnamefont {Zverev}}, \bibinfo {author}
  {\bibfnamefont {A.}~\bibnamefont {Alfonsov}}, \bibinfo {author}
  {\bibfnamefont {V.}~\bibnamefont {Kataev}}, \bibinfo {author} {\bibfnamefont
  {B.}~\bibnamefont {B\"{u}chner}}, \bibinfo {author} {\bibfnamefont {E.~F.}\
  \bibnamefont {Schwier}}, \bibinfo {author} {\bibfnamefont {S.}~\bibnamefont
  {Kumar}}, \bibinfo {author} {\bibfnamefont {A.}~\bibnamefont {Kimura}},
  \bibinfo {author} {\bibfnamefont {L.}~\bibnamefont {Petaccia}}, \bibinfo
  {author} {\bibfnamefont {G.}~\bibnamefont {Di~Santo}}, \bibinfo {author}
  {\bibfnamefont {R.~C.}\ \bibnamefont {Vidal}}, \bibinfo {author}
  {\bibfnamefont {S.}~\bibnamefont {Schatz}}, \bibinfo {author} {\bibfnamefont
  {K.}~\bibnamefont {Kißner}}, \bibinfo {author} {\bibfnamefont
  {M.}~\bibnamefont {\"{U}nzelmann}}, \bibinfo {author} {\bibfnamefont {C.~H.}\
  \bibnamefont {Min}}, \bibinfo {author} {\bibfnamefont {S.}~\bibnamefont
  {Moser}}, \bibinfo {author} {\bibfnamefont {T.~R.~F.}\ \bibnamefont
  {Peixoto}}, \bibinfo {author} {\bibfnamefont {F.}~\bibnamefont {Reinert}},
  \bibinfo {author} {\bibfnamefont {A.}~\bibnamefont {Ernst}}, \bibinfo
  {author} {\bibfnamefont {P.~M.}\ \bibnamefont {Echenique}}, \bibinfo {author}
  {\bibfnamefont {A.}~\bibnamefont {Isaeva}},\ and\ \bibinfo {author}
  {\bibfnamefont {E.~V.}\ \bibnamefont {Chulkov}},\ }\bibfield  {title}
  {\bibinfo {title} {Prediction and observation of an antiferromagnetic
  topological insulator},\ }\href {https://doi.org/10.1038/s41586-019-1840-9}
  {\bibfield  {journal} {\bibinfo  {journal} {Nature}\ }\textbf {\bibinfo
  {volume} {576}},\ \bibinfo {pages} {416–422} (\bibinfo {year}
  {2019})}\BibitemShut {NoStop}%
\bibitem [{\citenamefont {Tang}\ \emph {et~al.}(2016)\citenamefont {Tang},
  \citenamefont {Zhou}, \citenamefont {Xu},\ and\ \citenamefont
  {Zhang}}]{tang2016dirac}%
  \BibitemOpen
  \bibfield  {author} {\bibinfo {author} {\bibfnamefont {P.}~\bibnamefont
  {Tang}}, \bibinfo {author} {\bibfnamefont {Q.}~\bibnamefont {Zhou}}, \bibinfo
  {author} {\bibfnamefont {G.}~\bibnamefont {Xu}},\ and\ \bibinfo {author}
  {\bibfnamefont {S.-C.}\ \bibnamefont {Zhang}},\ }\bibfield  {title} {\bibinfo
  {title} {Dirac fermions in an antiferromagnetic semimetal},\ }\href
  {https://doi.org/10.1038/nphys3839} {\bibfield  {journal} {\bibinfo
  {journal} {Nat. Phys.}\ }\textbf {\bibinfo {volume} {12}},\ \bibinfo {pages}
  {1100–1104} (\bibinfo {year} {2016})}\BibitemShut {NoStop}%
\bibitem [{\citenamefont {Khusyainov}\ \emph {et~al.}(2023)\citenamefont
  {Khusyainov}, \citenamefont {Gareev}, \citenamefont {Radovskaia},
  \citenamefont {Sampathkumar}, \citenamefont {Acharya}, \citenamefont
  {Šiškins}, \citenamefont {Mañas-Valero}, \citenamefont {Ivanov},
  \citenamefont {Coronado}, \citenamefont {Rasing}, \citenamefont {Kimel},\
  and\ \citenamefont {Afanasiev}}]{khusyainov2023ultrafast}%
  \BibitemOpen
  \bibfield  {author} {\bibinfo {author} {\bibfnamefont {D.}~\bibnamefont
  {Khusyainov}}, \bibinfo {author} {\bibfnamefont {T.}~\bibnamefont {Gareev}},
  \bibinfo {author} {\bibfnamefont {V.}~\bibnamefont {Radovskaia}}, \bibinfo
  {author} {\bibfnamefont {K.}~\bibnamefont {Sampathkumar}}, \bibinfo {author}
  {\bibfnamefont {S.}~\bibnamefont {Acharya}}, \bibinfo {author} {\bibfnamefont
  {M.}~\bibnamefont {Šiškins}}, \bibinfo {author} {\bibfnamefont
  {S.}~\bibnamefont {Mañas-Valero}}, \bibinfo {author} {\bibfnamefont {B.~A.}\
  \bibnamefont {Ivanov}}, \bibinfo {author} {\bibfnamefont {E.}~\bibnamefont
  {Coronado}}, \bibinfo {author} {\bibfnamefont {T.}~\bibnamefont {Rasing}},
  \bibinfo {author} {\bibfnamefont {A.~V.}\ \bibnamefont {Kimel}},\ and\
  \bibinfo {author} {\bibfnamefont {D.}~\bibnamefont {Afanasiev}},\ }\bibfield
  {title} {\bibinfo {title} {{Ultrafast laser-induced spin–lattice dynamics
  in the van der Waals antiferromagnet CoPS3}},\ }\bibfield  {journal}
  {\bibinfo  {journal} {APL Mater.}\ }\textbf {\bibinfo {volume} {11}},\ \href
  {https://doi.org/10.1063/5.0146128} {10.1063/5.0146128} (\bibinfo {year}
  {2023})\BibitemShut {NoStop}%
\bibitem [{\citenamefont {Lee}\ \emph {et~al.}(2022)\citenamefont {Lee},
  \citenamefont {Son}, \citenamefont {Kim}, \citenamefont {Kang}, \citenamefont
  {Shen}, \citenamefont {Kenzelmann}, \citenamefont {Delley}, \citenamefont
  {Savchenko}, \citenamefont {Parchenko}, \citenamefont {Na}, \citenamefont
  {Choi}, \citenamefont {Kim}, \citenamefont {Cheong}, \citenamefont {Derlet},
  \citenamefont {Kleibert},\ and\ \citenamefont {Park}}]{lee2023giant}%
  \BibitemOpen
  \bibfield  {author} {\bibinfo {author} {\bibfnamefont {Y.}~\bibnamefont
  {Lee}}, \bibinfo {author} {\bibfnamefont {S.}~\bibnamefont {Son}}, \bibinfo
  {author} {\bibfnamefont {C.}~\bibnamefont {Kim}}, \bibinfo {author}
  {\bibfnamefont {S.}~\bibnamefont {Kang}}, \bibinfo {author} {\bibfnamefont
  {J.}~\bibnamefont {Shen}}, \bibinfo {author} {\bibfnamefont {M.}~\bibnamefont
  {Kenzelmann}}, \bibinfo {author} {\bibfnamefont {B.}~\bibnamefont {Delley}},
  \bibinfo {author} {\bibfnamefont {T.}~\bibnamefont {Savchenko}}, \bibinfo
  {author} {\bibfnamefont {S.}~\bibnamefont {Parchenko}}, \bibinfo {author}
  {\bibfnamefont {W.}~\bibnamefont {Na}}, \bibinfo {author} {\bibfnamefont
  {K.}~\bibnamefont {Choi}}, \bibinfo {author} {\bibfnamefont {W.}~\bibnamefont
  {Kim}}, \bibinfo {author} {\bibfnamefont {H.}~\bibnamefont {Cheong}},
  \bibinfo {author} {\bibfnamefont {P.~M.}\ \bibnamefont {Derlet}}, \bibinfo
  {author} {\bibfnamefont {A.}~\bibnamefont {Kleibert}},\ and\ \bibinfo
  {author} {\bibfnamefont {J.}~\bibnamefont {Park}},\ }\bibfield  {title}
  {\bibinfo {title} {{Giant Magnetic Anisotropy in the Atomically Thin van der
  Waals Antiferromagnet FePS$_3$}},\ }\bibfield  {journal} {\bibinfo  {journal}
  {Adv. Electron. Mater.}\ }\textbf {\bibinfo {volume} {9}},\ \href
  {https://doi.org/10.1002/aelm.202200650} {10.1002/aelm.202200650} (\bibinfo
  {year} {2022})\BibitemShut {NoStop}%
\bibitem [{\citenamefont {Dhakal}\ \emph {et~al.}(2024)\citenamefont {Dhakal},
  \citenamefont {Griffith},\ and\ \citenamefont {Winter}}]{dhakal2024hybrid}%
  \BibitemOpen
  \bibfield  {author} {\bibinfo {author} {\bibfnamefont {R.}~\bibnamefont
  {Dhakal}}, \bibinfo {author} {\bibfnamefont {S.}~\bibnamefont {Griffith}},\
  and\ \bibinfo {author} {\bibfnamefont {S.~M.}\ \bibnamefont {Winter}},\
  }\bibfield  {title} {\bibinfo {title} {{Hybrid spin-orbit exciton-magnon
  excitations in FePS3}},\ }\bibfield  {journal} {\bibinfo  {journal} {npj
  Quantum Mater.}\ }\textbf {\bibinfo {volume} {9}},\ \href
  {https://doi.org/10.1038/s41535-024-00675-x} {10.1038/s41535-024-00675-x}
  (\bibinfo {year} {2024})\BibitemShut {NoStop}%
\bibitem [{\citenamefont {Tatara}(2019)}]{tatara2019effective}%
  \BibitemOpen
  \bibfield  {author} {\bibinfo {author} {\bibfnamefont {G.}~\bibnamefont
  {Tatara}},\ }\bibfield  {title} {\bibinfo {title} {Effective gauge field
  theory of spintronics},\ }\href
  {https://doi.org/https://doi.org/10.1016/j.physe.2018.05.011} {\bibfield
  {journal} {\bibinfo  {journal} {Physica E: Low-dimensional Systems and
  Nanostructures}\ }\textbf {\bibinfo {volume} {106}},\ \bibinfo {pages} {208}
  (\bibinfo {year} {2019})}\BibitemShut {NoStop}%
\bibitem [{\citenamefont {Tang}\ and\ \citenamefont
  {Cheng}(2024)}]{tang2024lossless}%
  \BibitemOpen
  \bibfield  {author} {\bibinfo {author} {\bibfnamefont {J.}~\bibnamefont
  {Tang}}\ and\ \bibinfo {author} {\bibfnamefont {R.}~\bibnamefont {Cheng}},\
  }\bibfield  {title} {\bibinfo {title} {{Lossless Spin-Orbit Torque in
  Antiferromagnetic Topological Insulator
  ${\mathrm{MnBi}}_{2}{\mathrm{Te}}_{4}$}},\ }\href
  {https://doi.org/10.1103/PhysRevLett.132.136701} {\bibfield  {journal}
  {\bibinfo  {journal} {Phys. Rev. Lett.}\ }\textbf {\bibinfo {volume} {132}},\
  \bibinfo {pages} {136701} (\bibinfo {year} {2024})}\BibitemShut {NoStop}%
\bibitem [{\citenamefont {McCreary}\ \emph {et~al.}(2020)\citenamefont
  {McCreary}, \citenamefont {Simpson}, \citenamefont {Mai}, \citenamefont
  {McMichael}, \citenamefont {Douglas}, \citenamefont {Butch}, \citenamefont
  {Dennis}, \citenamefont {Vald\'es~Aguilar},\ and\ \citenamefont
  {Hight~Walker}}]{McCreary2020}%
  \BibitemOpen
  \bibfield  {author} {\bibinfo {author} {\bibfnamefont {A.}~\bibnamefont
  {McCreary}}, \bibinfo {author} {\bibfnamefont {J.~R.}\ \bibnamefont
  {Simpson}}, \bibinfo {author} {\bibfnamefont {T.~T.}\ \bibnamefont {Mai}},
  \bibinfo {author} {\bibfnamefont {R.~D.}\ \bibnamefont {McMichael}}, \bibinfo
  {author} {\bibfnamefont {J.~E.}\ \bibnamefont {Douglas}}, \bibinfo {author}
  {\bibfnamefont {N.}~\bibnamefont {Butch}}, \bibinfo {author} {\bibfnamefont
  {C.}~\bibnamefont {Dennis}}, \bibinfo {author} {\bibfnamefont
  {R.}~\bibnamefont {Vald\'es~Aguilar}},\ and\ \bibinfo {author} {\bibfnamefont
  {A.~R.}\ \bibnamefont {Hight~Walker}},\ }\bibfield  {title} {\bibinfo {title}
  {Quasi-two-dimensional magnon identification in antiferromagnetic ${\rm
  feps}_3$ via magneto-raman spectroscopy},\ }\href
  {https://doi.org/10.1103/PhysRevB.101.064416} {\bibfield  {journal} {\bibinfo
   {journal} {Phys. Rev. B}\ }\textbf {\bibinfo {volume} {101}},\ \bibinfo
  {pages} {064416} (\bibinfo {year} {2020})}\BibitemShut {NoStop}%
\bibitem [{bla(2023)}]{blank2023two}%
  \BibitemOpen
  \bibfield  {title} {\bibinfo {title} {{Two-Dimensional Terahertz Spectroscopy
  of Nonlinear Phononics in the Topological Insulator
  ${\mathrm{MnBi}}_{2}{\mathrm{Te}}_{4}$}, author = {Blank, T. G. H. and
  Grishunin, K. A. and Zvezdin, K. A. and Hai, N. T. and Wu, J. C. and Su,
  S.-H. and Huang, J.-C. A. and Zvezdin, A. K. and Kimel, A. V.}},\ }\href
  {https://doi.org/10.1103/PhysRevLett.131.026902} {\bibfield  {journal}
  {\bibinfo  {journal} {Phys. Rev. Lett.}\ }\textbf {\bibinfo {volume} {131}},\
  \bibinfo {pages} {026902} (\bibinfo {year} {2023})}\BibitemShut {NoStop}%
\end{thebibliography}%

\clearpage

\section*{End Matter}

\appendix
\section{AFCP induced back action on N\'eel order}
\label{app:afcp_control}

To analyze the back-action of an AFCP on antiferromagnetic order, we couple the two-mode phonon model to the N\'eel-vector dynamics described by the Lagrangian
\begin{align}
    L=& \frac{1}{2}\dot{\bm Q}^{2}
    -\frac{1}{2}\bm Q^{\rm T}D_0\bm Q
    +g N_z L_{\rm N}
    +Z_{\rm IR}E(t)Q_{\rm IR} \\
    +& \sum_{\alpha = A,B}(1-\cos\theta_\alpha)\dot\phi_\alpha
    - E_{\rm AFM}(\theta_A,\phi_A,\theta_B,\phi_B) \notag
\end{align}
where $Z_{\rm IR}$ is the Born effective charge of the IR phonon and $E(t)$ is the THz electric field parallel to the electric dipole moment of the IR phonon. The angles $\theta_\alpha$ and $\phi_\alpha$ specify the orientation of the unit spin vector $\bm s_\alpha$ on sublattice $\alpha=A,B$. The last two terms are the spin Berry phase and the antiferromagnetic energy, respectively, which generate the Landau-Lifshitz spin-precession dynamics~\cite{tatara2019effective, tang2024lossless}. We take
\begin{equation}
    E_{\rm AFM}
    =H_E\bm s_A\cdot\bm s_B
    -\frac{H_A}{2}\left(s_{A,z}^2+s_{B,z}^2\right)
    \label{eq:em_hmag}
\end{equation}
where $H_E>0$ is the antiferromagnetic exchange and $H_A>0$ is the easy-axis anisotropy. The normalized N\'eel vector and net magnetization are $\bm n=(\bm s_A-\bm s_B)/2$ and $\bm m=(\bm s_A+\bm s_B)/2$, respectively.

In the two-mode subspace, $\bm Q=(Q_{\rm R},Q_{\rm IR})^{\rm T}$, $D_0=\operatorname{diag}(\omega_{\rm R}^2,\omega_{\rm IR}^2)$, and the mode-space angular momentum is
\begin{equation}
    \ell_Q=Q_{\rm R}\dot Q_{\rm IR}-Q_{\rm IR}\dot Q_{\rm R}
    =(\mathsf J\bm Q)\cdot\dot{\bm Q},
    \qquad
    \mathsf J=
    \begin{pmatrix}
        0&-1\\
        1&0
    \end{pmatrix}. \notag
    \label{eq:em_lq}
\end{equation}
Here we take $N_z=n_z$, $L_{\rm N} = \ell_Q$, and the corresponding coupling strength is $g=\Omega_0/2$. It is noted that $g$ and $\Omega_0$ here are independent of the N\'eel vector orientation. The dependence has been captured by the $n_z$ factor already.

Variation with respect to $\bm Q$ and adding a phenomenological phonon damping rate $\gamma_{\rm ph}$  give
\begin{align}
    \ddot{\bm Q}+\gamma_{\rm ph}\dot{\bm Q}+D_0\bm Q
    &+\Omega_0n_z\mathsf J\dot{\bm Q}
    +\frac{\Omega_0}{2}\dot n_z\mathsf J\bm Q \notag\\
    &=Z_{\rm IR}E(t)\,(0,1)^{\rm T},
    \label{eq:em_q}
\end{align}
where both $n_z$ and its dynamics influence the phonon motion. 
Phonons reciprocally act on the spins dynamics, described by the equations of motion \begin{equation}
    \begin{split}
    \dot{\bm s}_{\alpha}=-\frac{1}{1+\alpha_G^2}
    \bigl[&\bm s_{\alpha}\times\bm H_{\alpha}^{\rm eff}+\alpha_G\bm s_{\alpha}\times
      (\bm s_{\alpha}\times\bm H_{\alpha}^{\rm eff})\bigr]
    \end{split}
    \label{eq:em_llg}
\end{equation}
where $\alpha_G$ is the Gilbert damping parameter, and the effective fields are
\begin{align}
    \bm H_A^{\rm eff}
    &=-H_E\bm s_B+H_As_{A,z}\hat z
      +\frac{g}{2}L_{\rm N}\hat z,\notag\\
    \bm H_B^{\rm eff}
    &=-H_E\bm s_A+H_As_{B,z}\hat z
      -\frac{g}{2}L_{\rm N}\hat z.
    \label{eq:em_fields}
\end{align}
Thus, a coherent AFCP generates the staggered effective field $\bm h_{\rm ph}(t)=\frac{g}{2}L_{\rm N}(t)\hat z$,
which enters with opposite signs on the two sublattices and therefore couples to $\bm n$. 

\begin{figure}[t]
    \centering
    \includegraphics[width=0.5 \textwidth]{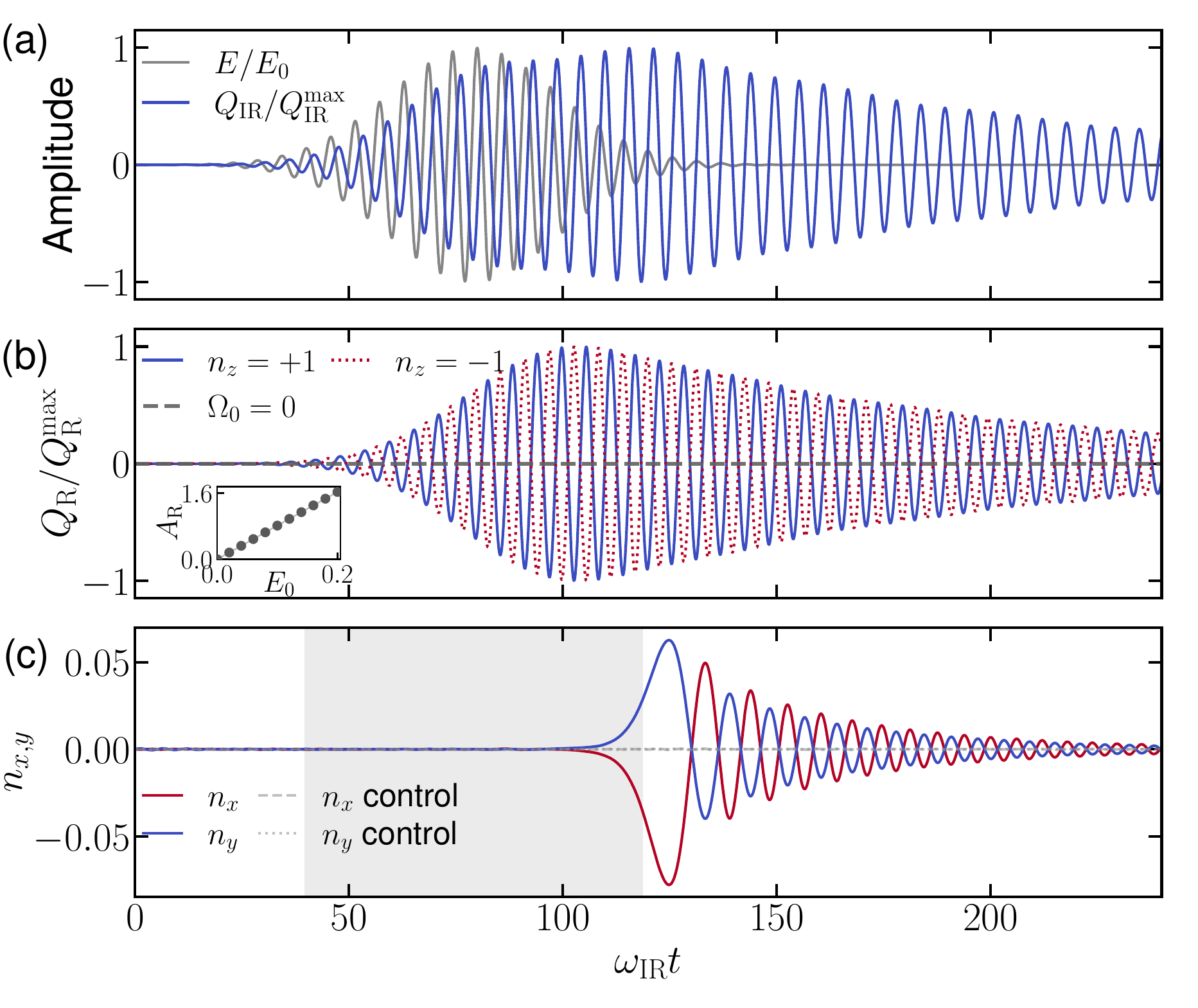}
    \caption{THz light-driven spin-phonon dynamics. (a) Normalized Gaussian-enveloped THz field and the driven IR coordinate. (b) Normalized Raman coordinate for the two opposite N\'eel vectors. Reversing $n_z$ reverses $Q_{\rm R}$, whereas $Q_{\rm R}$ vanishes for $\Omega_0=0$. The inset shows the peak Raman amplitude $A_{\rm R}=\max |Q_{\rm R}|$ vs $E_0$. (c) Transverse N\'eel components from the coupled dynamics (solid colored curves). The gray dashed and dotted curves are a same-noise control with phononic conjugate field being omitted. The shaded interval denotes $t_0-2\tau<t<t_0+2\tau$.} 
    \label{fig:afcp_back_action}
\end{figure}

To illustrate this back-action, we drive the IR mode with a Gaussian-enveloped THz pulse with central frequency $\omega_d$ and width $\tau$ as shown in Fig.~\ref{fig:afcp_back_action}(a):
\begin{align}
    E(t)=E_0\exp\!\left[-\frac{(t-t_0)^2}{2\tau^2}\right]
    \cos(\omega_dt),\qquad t_0=4\tau .
    \label{eq:em_pulse}
\end{align}
We solve the driven dynamics of the coupled spin-phonon system where stochastic fields $\bm\xi_\alpha(t)$ from thermal fluctuations are added to both spins with
\begin{equation}
    \left\langle\xi_{\alpha i}(t)\right\rangle=0,\qquad
    \left\langle\xi_{\alpha i}(t)\xi_{\beta j}(t')\right\rangle
    =2D\,\delta_{\alpha\beta}\delta_{ij}\delta(t-t') \notag
    \label{eq:em_noise}
\end{equation}
where $i,j=x,y,z$ and $D$ is the noise strength. In the simulation, a small phenomenological fluctuation $D=10^{-10}$ is used.

We use dimensionless units with $\omega_{\rm IR}=1$. The other parameters are set as $\omega_{\rm R}=1.05$, $\gamma_{\rm ph}=0.02$, $H_E=2.302$, $H_A=0.2736$, $\alpha_G=0.01$, $\omega_d=1.10$, and $\tau=19.8$. These parameters are motivated by the phonon and magnetic properties of FePS$_3$~\cite{McCreary2020,Kargar2020}. The resulting magnetic resonance frequency is $\omega_{\rm AF}=\sqrt{(2H_E+H_A)H_A}=1.155$. The gyroscopic coupling $\Omega_0$ is a control parameter in the model system. We take $\Omega_0=0.148$, Born effective charge $Z_{\rm IR}=1$, and $E_0=0.418$ as a minimal illustration of the N\'eel order's effect on the driven phonon dynamics and the back action. 

Figure~\ref{fig:afcp_back_action}(a) shows the response of the IR mode to the applied pulse, which further excites the Raman mode through the gyroscopic coupling. As shown in Fig.~\ref{fig:afcp_back_action}(b), the amplitude of Raman mode is zero when the gyroscopic coupling is turned off. Furthermore, the Raman response flips sign while the IR response remains unchanged as the N\'eel vector reverses. The peak Raman amplitude $A_{\rm R}$ scales linearly with $E_0$ in the weak-drive regime as shown in the inset. By contrast, the nonlinear phononics induced Raman response via couplings like $Q_{\rm IR}^2 Q_{\rm R}$ is insensitive to the N\'eel vector orientation and is quadratic in $E_0$~\cite{blank2023two}. 
These signatures allow one to use terahertz spectroscopy to probe the AFCP and its conjugate-field coupling to the N\'eel order in pump-probe experiments~\cite{blank2023two}. 

The presence of both Raman and IR modes induce nonzero sublattice staggered angular momentum $L_{\rm N}$, acting on the spins as a staggered field. When this field overcomes the anisotropic field, the fluctuations can be amplified into a coherent precession, as shown in Fig.~\ref{fig:afcp_back_action}(c). As a comparison, when the back action to the spins is omitted while the same stochastic fields are applied, there is no coherent spin dynamics as shown by the control lines. 




\end{document}